\documentclass[longauth]{aa} 

\usepackage{txfonts}
\usepackage{euclid}
\bibpunct{(}{)}{;}{a}{}{,} 

\usepackage{hyperref}
\usepackage{graphicx}
\usepackage{color} 		
\usepackage[all]{hypcap} 

\begin{document}

\newcommand{\MOPEX}{\texttt{MOPEX}}
\newcommand{\mum}{\,\si{\micron}\xspace}

\title{\Euclid preparation: XVIII. Cosmic Dawn Survey. Spitzer observations of the \Euclid deep fields and calibration fields}

\author{Euclid Collaboration: Andrea Moneti$^{1}$\thanks{\email{moneti@iap.fr}}, H.J.~McCracken$^{1,2}$, M.~Shuntov$^{1}$, O.B.~Kauffmann$^{3}$, P.~Capak$^{4}$, I.~Davidzon$^{4}$, O.~Ilbert$^{3}$, C.~Scarlata$^{5}$, S.~Toft$^{6,7}$, J.~Weaver$^{4}$
% ...
}

\institute{$^{1}$ Institut d'Astrophysique de Paris, 98bis Boulevard Arago, F-75014, Paris, France\\
$^{2}$ Sorbonne Universit{\'e}s, UPMC Univ Paris 6 et CNRS, UMR 7095, Institut d'Astrophysique de Paris, 98 bis bd Arago, 75014 Paris, France\\
$^{3}$ Aix-Marseille Univ, CNRS, CNES, LAM, Marseille, France\\
$^{4}$ Cosmic Dawn Center (DAWN), Niels Bohr Institute, University of Copenhagen, Vibenshuset, Lyngbyvej 2, DK-2100 Copenhagen, Denmark\\
$^{5}$ Minnesota Institute for Astrophysics, University of Minnesota, 116 Church St SE, Minneapolis, MN 55455, USA\\
$^{6}$ Cosmic Dawn Center (DAWN)\\
$^{7}$ Niels Bohr Institute, University of Copenhagen, Jagtvej 128, 2200 Copenhagen, Denmark\\
$^{8}$ Infrared Processing and Analysis Center, California Institute of Technology, Pasadena, CA 91125, USA\\
$^{9}$ California institute of Technology, 1200 E California Blvd, Pasadena, CA 91125, USA\\
$^{10}$ Physics and Astronomy Department, University of California, 900 University Ave., Riverside, CA 92521, USA\\
$^{11}$ Institute for Astronomy, University of Hawaii, 2680 Woodlawn Drive, Honolulu, HI 96822, USA\\
$^{12}$ INFN-Sezione di Roma, Piazzale Aldo Moro, 2 - c/o Dipartimento di Fisica, Edificio G. Marconi, I-00185 Roma, Italy\\
$^{13}$ INAF-Osservatorio Astronomico di Roma, Via Frascati 33, I-00078 Monteporzio Catone, Italy\\
$^{14}$ Jet Propulsion Laboratory, California Institute of Technology, 4800 Oak Grove Drive, Pasadena, CA, 91109, USA\\
$^{15}$ Institute of Cosmology and Gravitation, University of Portsmouth, Portsmouth PO1 3FX, UK\\
$^{16}$ INAF-Osservatorio di Astrofisica e Scienza dello Spazio di Bologna, Via Piero Gobetti 93/3, I-40129 Bologna, Italy\\
$^{17}$ Max Planck Institute for Extraterrestrial Physics, Giessenbachstr. 1, D-85748 Garching, Germany\\
$^{18}$ INAF-Osservatorio Astrofisico di Torino, Via Osservatorio 20, I-10025 Pino Torinese (TO), Italy\\
$^{19}$ INFN-Sezione di Roma Tre, Via della Vasca Navale 84, I-00146, Roma, Italy\\
$^{20}$ Department of Mathematics and Physics, Roma Tre University, Via della Vasca Navale 84, I-00146 Rome, Italy\\
$^{21}$ Institut de Recherche en Astrophysique et Plan\'etologie (IRAP), Universit\'e de Toulouse, CNRS, UPS, CNES, 14 Av. Edouard Belin, F-31400 Toulouse, France\\
$^{22}$ INAF-Osservatorio Astronomico di Capodimonte, Via Moiariello 16, I-80131 Napoli, Italy\\
$^{23}$ Centro de Astrof\'{\i}sica da Universidade do Porto, Rua das Estrelas, 4150-762 Porto, Portugal\\
$^{24}$ Instituto de Astrof\'isica e Ci\^encias do Espa\c{c}o, Universidade do Porto, CAUP, Rua das Estrelas, PT4150-762 Porto, Portugal\\
$^{25}$ INAF-IASF Milano, Via Alfonso Corti 12, I-20133 Milano, Italy\\
$^{26}$ Port d'Informaci\'{o} Cient\'{i}fica, Campus UAB, C. Albareda s/n, 08193 Bellaterra (Barcelona), Spain\\
$^{27}$ Institut de F\'{i}sica d’Altes Energies (IFAE), The Barcelona Institute of Science and Technology, Campus UAB, 08193 Bellaterra (Barcelona), Spain\\
$^{28}$ Institute of Space Sciences (ICE, CSIC), Campus UAB, Carrer de Can Magrans, s/n, 08193 Barcelona, Spain\\
$^{29}$ Institut d’Estudis Espacials de Catalunya (IEEC), Carrer Gran Capit\'a 2-4, 08034 Barcelona, Spain\\
$^{30}$ Department of Physics "E. Pancini", University Federico II, Via Cinthia 6, I-80126, Napoli, Italy\\
$^{31}$ INFN section of Naples, Via Cinthia 6, I-80126, Napoli, Italy\\
$^{32}$ Dipartimento di Fisica e Astronomia ''Augusto Righi'' - Alma Mater Studiorum Universit\'a di Bologna, Viale Berti Pichat 6/2, I-40127 Bologna, Italy\\
$^{33}$ INAF-Osservatorio Astrofisico di Arcetri, Largo E. Fermi 5, I-50125, Firenze, Italy\\
$^{34}$ Institut national de physique nucl\'eaire et de physique des particules, 3 rue Michel-Ange, 75794 Paris C\'edex 16, France\\
$^{35}$ Centre National d'Etudes Spatiales, Toulouse, France\\
$^{36}$ Institute for Astronomy, University of Edinburgh, Royal Observatory, Blackford Hill, Edinburgh EH9 3HJ, UK\\
$^{37}$ Jodrell Bank Centre for Astrophysics, School of Physics and Astronomy, University of Manchester, Oxford Road, Manchester M13 9PL, UK\\
$^{38}$ European Space Agency/ESRIN, Largo Galileo Galilei 1, 00044 Frascati, Roma, Italy\\
$^{39}$ ESAC/ESA, Camino Bajo del Castillo, s/n., Urb. Villafranca del Castillo, 28692 Villanueva de la Ca\~nada, Madrid, Spain\\
$^{40}$ Univ Lyon, Univ Claude Bernard Lyon 1, CNRS/IN2P3, IP2I Lyon, UMR 5822, F-69622, Villeurbanne, France\\
$^{41}$ Mullard Space Science Laboratory, University College London, Holmbury St Mary, Dorking, Surrey RH5 6NT, UK\\
$^{42}$ Departamento de F\'isica, Faculdade de Ci\^encias, Universidade de Lisboa, Edif\'icio C8, Campo Grande, PT1749-016 Lisboa, Portugal\\
$^{43}$ Instituto de Astrof\'isica e Ci\^encias do Espa\c{c}o, Faculdade de Ci\^encias, Universidade de Lisboa, Campo Grande, PT-1749-016 Lisboa, Portugal\\
$^{44}$ Department of Astronomy, University of Geneva, ch. d\'Ecogia 16, CH-1290 Versoix, Switzerland\\
$^{45}$ Universit\'e Paris-Saclay, CNRS, Institut d'astrophysique spatiale, 91405, Orsay, France\\
$^{46}$ Department of Physics, Oxford University, Keble Road, Oxford OX1 3RH, UK\\
$^{47}$ INFN-Padova, Via Marzolo 8, I-35131 Padova, Italy\\
$^{48}$ AIM, CEA, CNRS, Universit\'{e} Paris-Saclay, Universit\'{e} de Paris, F-91191 Gif-sur-Yvette, France\\
$^{49}$ INAF-Osservatorio Astronomico di Trieste, Via G. B. Tiepolo 11, I-34131 Trieste, Italy\\
$^{50}$ Istituto Nazionale di Astrofisica (INAF) - Osservatorio di Astrofisica e Scienza dello Spazio (OAS), Via Gobetti 93/3, I-40127 Bologna, Italy\\
$^{51}$ Istituto Nazionale di Fisica Nucleare, Sezione di Bologna, Via Irnerio 46, I-40126 Bologna, Italy\\
$^{52}$ INAF-Osservatorio Astronomico di Brera, Via Brera 28, I-20122 Milano, Italy\\
$^{53}$ INAF-Osservatorio Astronomico di Padova, Via dell'Osservatorio 5, I-35122 Padova, Italy\\
$^{54}$ Universit\"ats-Sternwarte M\"unchen, Fakult\"at f\"ur Physik, Ludwig-Maximilians-Universit\"at M\"unchen, Scheinerstrasse 1, 81679 M\"unchen, Germany\\
$^{55}$ Institute of Theoretical Astrophysics, University of Oslo, P.O. Box 1029 Blindern, N-0315 Oslo, Norway\\
$^{56}$ Leiden Observatory, Leiden University, Niels Bohrweg 2, 2333 CA Leiden, The Netherlands\\
$^{57}$ von Hoerner \& Sulger GmbH, Schlo{\ss}Platz 8, D-68723 Schwetzingen, Germany\\
$^{58}$ Max-Planck-Institut f\"ur Astronomie, K\"onigstuhl 17, D-69117 Heidelberg, Germany\\
$^{59}$ Aix-Marseille Univ, CNRS/IN2P3, CPPM, Marseille, France\\
$^{60}$ Universit\'e de Gen\`eve, D\'epartement de Physique Th\'eorique and Centre for Astroparticle Physics, 24 quai Ernest-Ansermet, CH-1211 Gen\`eve 4, Switzerland\\
$^{61}$ Department of Physics and Helsinki Institute of Physics, Gustaf H\"allstr\"omin katu 2, 00014 University of Helsinki, Finland\\
$^{62}$ NOVA optical infrared instrumentation group at ASTRON, Oude Hoogeveensedijk 4, 7991PD, Dwingeloo, The Netherlands\\
$^{63}$ Argelander-Institut f\"ur Astronomie, Universit\"at Bonn, Auf dem H\"ugel 71, 53121 Bonn, Germany\\
$^{64}$ Dipartimento di Fisica e Astronomia “Augusto Righi” - Alma Mater Studiorum Università di Bologna, via Piero Gobetti 93/2, I-40129 Bologna, Italy\\
$^{65}$ INFN-Sezione di Bologna, Viale Berti Pichat 6/2, I-40127 Bologna, Italy\\
$^{66}$ Centre for Extragalactic Astronomy, Department of Physics, Durham University, South Road, Durham, DH1 3LE, UK\\
$^{67}$ Universit\'e C\^{o}te d'Azur, Observatoire de la C\^{o}te d'Azur, CNRS, Laboratoire Lagrange, Bd de l'Observatoire, CS 34229, 06304 Nice cedex 4, France\\
$^{68}$ Institute of Physics, Laboratory of Astrophysics, Ecole Polytechnique F\'{e}d\'{e}rale de Lausanne (EPFL), Observatoire de Sauverny, 1290 Versoix, Switzerland\\
$^{69}$ European Space Agency/ESTEC, Keplerlaan 1, 2201 AZ Noordwijk, The Netherlands\\
$^{70}$ Department of Physics and Astronomy, University of Aarhus, Ny Munkegade 120, DK–8000 Aarhus C, Denmark\\
$^{71}$ Institute of Space Science, Bucharest, Ro-077125, Romania\\
$^{72}$ Departamento de Astrof\'{i}sica, Universidad de La Laguna, E-38206, La Laguna, Tenerife, Spain\\
$^{73}$ Instituto de Astrof\'isica de Canarias, Calle V\'ia L\'actea s/n, E-38204, San Crist\'obal de La Laguna, Tenerife, Spain\\
$^{74}$ Dipartimento di Fisica e Astronomia “G.Galilei", Universit\'a di Padova, Via Marzolo 8, I-35131 Padova, Italy\\
$^{75}$ Centro de Investigaciones Energ\'eticas, Medioambientales y Tecnol\'ogicas (CIEMAT), Avenida Complutense 40, 28040 Madrid, Spain\\
$^{76}$ Instituto de Astrof\'isica e Ci\^encias do Espa\c{c}o, Faculdade de Ci\^encias, Universidade de Lisboa, Tapada da Ajuda, PT-1349-018 Lisboa, Portugal\\
$^{77}$ Universidad Polit\'ecnica de Cartagena, Departamento de Electr\'onica y Tecnolog\'ia de Computadoras, 30202 Cartagena, Spain\\
$^{78}$ INFN-Sezione di Torino, Via P. Giuria 1, I-10125 Torino, Italy\\
$^{79}$ Dipartimento di Fisica, Universit\'a degli Studi di Torino, Via P. Giuria 1, I-10125 Torino, Italy\\
$^{80}$ Universit\'e de Paris, CNRS, Astroparticule et Cosmologie, F-75013 Paris, France\\
$^{81}$ Space Science Data Center, Italian Space Agency, via del Politecnico snc, 00133 Roma, Italy\\
$^{82}$ IFPU, Institute for Fundamental Physics of the Universe, via Beirut 2, 34151 Trieste, Italy\\
$^{83}$ SISSA, International School for Advanced Studies, Via Bonomea 265, I-34136 Trieste TS, Italy\\
$^{84}$ INFN, Sezione di Trieste, Via Valerio 2, I-34127 Trieste TS, Italy\\
$^{85}$ Institut de Physique Th\'eorique, CEA, CNRS, Universit\'e Paris-Saclay F-91191 Gif-sur-Yvette Cedex, France\\
$^{86}$ INFN-Bologna, Via Irnerio 46, I-40126 Bologna, Italy\\
$^{87}$ Dipartimento di Fisica e Scienze della Terra, Universit\'a degli Studi di Ferrara, Via Giuseppe Saragat 1, I-44122 Ferrara, Italy\\
$^{88}$ INAF, Istituto di Radioastronomia, Via Piero Gobetti 101, I-40129 Bologna, Italy\\
$^{89}$ University of Lyon, UCB Lyon 1, CNRS/IN2P3, IUF, IP2I Lyon, France\\
$^{90}$ INAF-Istituto di Astrofisica e Planetologia Spaziali, via del Fosso del Cavaliere, 100, I-00100 Roma, Italy\\
$^{91}$ INAF-IASF Bologna, Via Piero Gobetti 101, I-40129 Bologna, Italy\\
$^{92}$ School of Physics, HH Wills Physics Laboratory, University of Bristol, Tyndall Avenue, Bristol, BS8 1TL, UK\\
$^{93}$ Instituto de F\'isica Te\'orica UAM-CSIC, Campus de Cantoblanco, E-28049 Madrid, Spain\\
$^{94}$ Research Program in Systems Oncology, Faculty of Medicine, University of Helsinki, Helsinki, Finland\\
$^{95}$ Department of Physics, P.O. Box 64, 00014 University of Helsinki, Finland\\
$^{96}$ Department of Physics, Lancaster University, Lancaster, LA1 4YB, UK\\
$^{97}$ Department of Physics and Astronomy, University College London, Gower Street, London WC1E 6BT, UK\\
$^{98}$ Helsinki Institute of Physics, Gustaf H{\"a}llstr{\"o}min katu 2, University of Helsinki, Helsinki, Finland\\
$^{99}$ Centre de Calcul de l'IN2P3, 21 avenue Pierre de Coubertin F-69627 Villeurbanne Cedex, France\\
$^{100}$ Dipartimento di Fisica "Aldo Pontremoli", Universit\'a degli Studi di Milano, Via Celoria 16, I-20133 Milano, Italy\\
$^{101}$ INFN-Sezione di Milano, Via Celoria 16, I-20133 Milano, Italy\\
$^{102}$ Dipartimento di Fisica, Sapienza Universit\`a di Roma, Piazzale Aldo Moro 2, I-00185 Roma, Italy\\
$^{103}$ Institut f\"ur Theoretische Physik, University of Heidelberg, Philosophenweg 16, 69120 Heidelberg, Germany\\
$^{104}$ Zentrum f\"ur Astronomie, Universit\"at Heidelberg, Philosophenweg 12, D- 69120 Heidelberg, Germany\\
$^{105}$ INFN, Sezione di Lecce, Via per Arnesano, CP-193, I-73100, Lecce, Italy\\
$^{106}$ Department of Mathematics and Physics E. De Giorgi, University of Salento, Via per Arnesano, CP-I93, I-73100, Lecce, Italy\\
$^{107}$ Institute for Computational Science, University of Zurich, Winterthurerstrasse 190, 8057 Zurich, Switzerland\\
$^{108}$ Departamento de F\'isica, FCFM, Universidad de Chile, Blanco Encalada 2008, Santiago, Chile\\
$^{109}$ Department of Physics, P.O.Box 35 (YFL), 40014 University of Jyv\"askyl\"a, Finland\\
$^{110}$ Ruhr University Bochum, Faculty of Physics and Astronomy, Astronomical Institute (AIRUB), German Centre for Cosmological Lensing (GCCL), 44780 Bochum, Germany\\
}

\abstract{We present a new infrared survey covering the three \Euclid deep fields and  four other \Euclid calibration fields using \Spitzer's Infrared Array Camera (IRAC). We have combined these new observations with all relevant IRAC archival data of these fields in order to produce the deepest possible mosaics of these regions. In total, these observations represent nearly 11\,\% of the total \Spitzer mission time. The resulting mosaics cover a total of approximately 71.5\,deg$^2$ in the 3.6 and 4.5\mum bands, and approximately 21.8\,deg$^2$ in the 5.8 and 8\mum bands. They reach at least 24 AB magnitude (measured to $1\sigma$, in a $2\farcs5$ aperture) in the 3.6\mum band and up to $\sim 5$\,mag deeper in the deepest regions. The astrometry is tied to the \Gaia astrometric reference system, and the typical astrometric uncertainty for sources with $16<[3.6]<19$ is $\lesssim 0\farcs15$. The photometric calibration is in excellent agreement with previous  WISE measurements. We have extracted source number counts from the 3.6\mum band mosaics and they are in excellent agreement with previous measurements. Given that the \SpitzerST has now been decommissioned these mosaics are likely to be the definitive reduction of these IRAC data. This survey therefore represents an essential first step in assembling multi-wavelength data on the \Euclid deep fields which are set to become some of the premier fields for extragalactic astronomy in the 2020s.}


\titlerunning{The Cosmic Dawn Survey}
\authorrunning{Moneti et al.}
\date{Released on TBD / Accepted date TBD}

\maketitle
\section{Introduction} 
\label{sec:intro}

The \Euclid mission will survey 15\,000\,$\deg^2$ of the extragalactic sky to investigate the nature of dark energy and dark matter, and to study the formation and evolution of galaxies \citep{laureijs_euclid_2011}. To this end \Euclid will obtain high resolution and high signal-to-noise imaging of a billion galaxies in a broad optical filter to measure their shapes and in three near-infrared (NIR) filters to measure their colours. It will also obtain high signal-to-noise NIR spectroscopy of about thirty million of these galaxies to measure abundances and redshifts. Additionally, photometric redshifts will be determined by combining the \Euclid data with optical photometry from external surveys. 

To reach the required precision on cosmological parameters and satisfy the stringent mission requirements on completeness, spectroscopic purity and shape noise bias, \Euclid must also obtain observations with 40 times longer exposure per pixel than the main survey over regions covering at least 40\,deg$^2$. To this end, three `deep'  fields have been selected by the Euclid Consortium. They are described in detail in \cite{2021arXiv210801201S} and we give just a very brief description here. They are: (1) the Euclid Deep Field North (EDF-N), a roughly circular, 10\,deg$^2$ region centred on the well-studied north Ecliptic pole, (2) the Euclid Deep Field Fornax (EDF-F), also roughly circular and of 10\,deg$^2$, centred on the Chandra Deep Field South and including the GOODS-S \citep{giavalisco_great_2004} and the Hubble Ultra Deep Field \citep{beckwith_hubble_2006}, and (3) the Euclid Deep Field South (EDF-S), a pill-shaped area of 20\,deg$^2$ with no previous dedicated observations. In addition to these three deep fields \Euclid will observe several fields for the calibration of photometric redshifts (photo-$z$). These fields need to be observed to a level 5 times deeper than the main survey and they are centred on some of the best studied extragalactic survey fields that already have extensive spectroscopic data: (1) the COSMOS field (2) the Extended Groth Strip (EGS), (3) the Hubble Deep Field North (HDF, also GOODS-N), and (4) the XMM-Large Scale Structure Survey field, which includes the Subaru XMM Deep Survey field (SXDS), and VIMOS VLT Deep Survey (VVDS)\footnote{These are now known as the "Euclid Auxiliary Fields" in \Euclid terminology}.

While \Euclid\ will observe these fields primarily for calibration purposes, those observations will provide an unprecedented data set to study galaxies to faint magnitudes and high redshifts. The survey efficiency of \Euclid in the NIR bands is orders of magnitudes greater than that of ground-based telescopes (e.g., VISTA). The \Euclid deep fields alone will be 30 times larger and one magnitude deeper than the latest UltraVISTA data release covering the COSMOS field, and will reach a depth of 26\,mag in the $Y$, $J$, $H$ filters ($5\,\sigma$). In addition, \Euclid carries a wide-field near-infrared grism spectrograph, the Near Infrared Spectrometer and Photometer (NISP), covering the $0.92 < \lambda < 1.85\mum$ region, which will provide multiple spectra at numerous grism orientations for more than one million sources to a line flux limit similar to 3D-HST \citep{Brammer12_3DHST} and over an area 200 times larger than the COSMOS field (depending on the scheduling of the blue grism observations). The observations of the deep fields will result in the most complete and deepest spectroscopic coverage produced by \Euclid. Such a spectroscopic data set will be unique for the reconstruction of the galaxy environment at cosmic noon and for measuring the star formation rate from the H$\alpha$ emission line intensity.

The deep and wide NIR data from \Euclid are also ideal for detecting significant numbers of high-redshift ($7<z<10$) galaxies, as the Lyman $\alpha$ line is redshifted out of the optical into the NIR. However, in order to distinguish galaxy candidates from stars (primarily brown dwarfs), faint Balmer-break galaxies, and dusty star-forming galaxies at lower redshifts which all can have similar NIR magnitudes and colours, deep optical and mid-infrared (MIR) data are also needed \citep[]{bouwens2019,bridge19,Bowler20_LF}. 

The Cosmic Dawn Survey (Toft et al., in prep) aims to obtain uniform, multi-wavelength imaging of the \Euclid deep and calibration fields to limits matching the Euclid data for characterisation of high-redshift galaxies. The optical data will be provided by the Hawaii-Two-0 Subaru telescope/Hyper-SuprimeCam (HSC) survey (McPartland et al., in prep.) for the EDF-N and EDF-F  and likely by the Vera C. Rubin Observatory for EDF-S and EDF-F. For the COSMOS and SXDS fields optical data are provided by the Subaru HSC Strategic program \citep[HSC-SSP][]{aihara2011}.   

In this paper, we present the \SpitzerST \citep{werner_spitzer_2004} component of the Cosmic Dawn Survey, consisting primarily of 3.6 and 4.5\mum observations of the three deep fields and parts of the calibration fields acquired with \Spitzer's IRAC camera \citep{fazio_infrared_2004}. Two dedicated programs were submitted for this purpose: the \Euclid/WFIRST Spitzer Legacy Survey (SLS, requesting 5\,286\,h, PI: Capak) covering the EDF-N and EDF-F fields, and the EDF-S survey (requesting 687\,h, PI: Scarlata).  These programmes were established based on the \Euclid plans for the deep fields that were available at the time. All the fields had been observed, at least in part, before our new observations, and we processed our new data together with all relevant archival IRAC data, thus including data obtained during the cryogenic mission, i.e., data at 5.8 and 8.0\mum. In this way we strive to produce the deepest possible MIR images (mosaics) of these fields to date. A significant improvement in our processing is that our pipeline ties the astrometry to the \Gaia reference system which, given its higher precision, will greatly facilitate cross-identification with other data, which will of course also have to be tied to \Gaia.

In addition to being essential for the identification of high-redshift galaxies, MIR data are crucial to reveal the stellar mass content of the high-redshift Universe (which is outside the scope of the \Euclid core science). The \Euclid data alone are not sufficient to characterise the stellar masses at $z>3.5$, as the Balmer break is redshifted out of the reddest band of the NISP. Without MIR data, the interpretation of spectral energy distributions (SEDs) would rely on rest-frame ultraviolet emission which is strongly affected by dust attenuation and dominated by stellar light of new-born stars. Therefore, integrated quantities like the stellar mass would be highly unreliable  \citep[][]{bell01}. Moreover, photometric redshifts would be prone to catastrophic failures resulting from the mis-identification of the Lyman and Balmer breaks \cite[e.g.][]{lefevre15, kauffmann20}. In summary, \Spitzer/IRAC data are crucial for identifying the most distant objects \citep[e.g.][]{bridge19}, for improving the accuracy of their photometric redshifts and for deriving their physical properties such as stellar masses, dust content, age, and star-formation rate from population synthesis models \citep[e.g.][]{perez-gonzalez08, caputi15, davidzon_cosmos2015_2017}. The build-up of stellar mass, especially when confronted with the amount of matter residing in dark matter halos at high redshifts can be a highly discriminating test for galaxy formation models \citep{legrand_cosmos-ultravista_2019}. Extrapolation of recent work in the COSMOS field \citep{Bowler20_LF} suggests that hundreds of the rarest, brightest $z>7$ galaxies are expected to be discovered in the \Euclid deep fields. These provide unique constraints on cosmic reionisation, as the brightest galaxies form in the highest density regions of the Universe which are expected to be the sites of the first generation of stars and galaxies, and thus of reionisation bubbles \citep{trac08}. 

The layout of this paper is as follows: Sect.~\ref{sec:obs} describes the observations, Sect.~\ref{sec:pro} presents our data processing techniques, and Sect.~\ref{sec:val} compares our results to previous ones. 

\section{Observations}
\label{sec:obs}

\begin{figure*}[hbt] 
    \includegraphics[width=0.95\hsize]{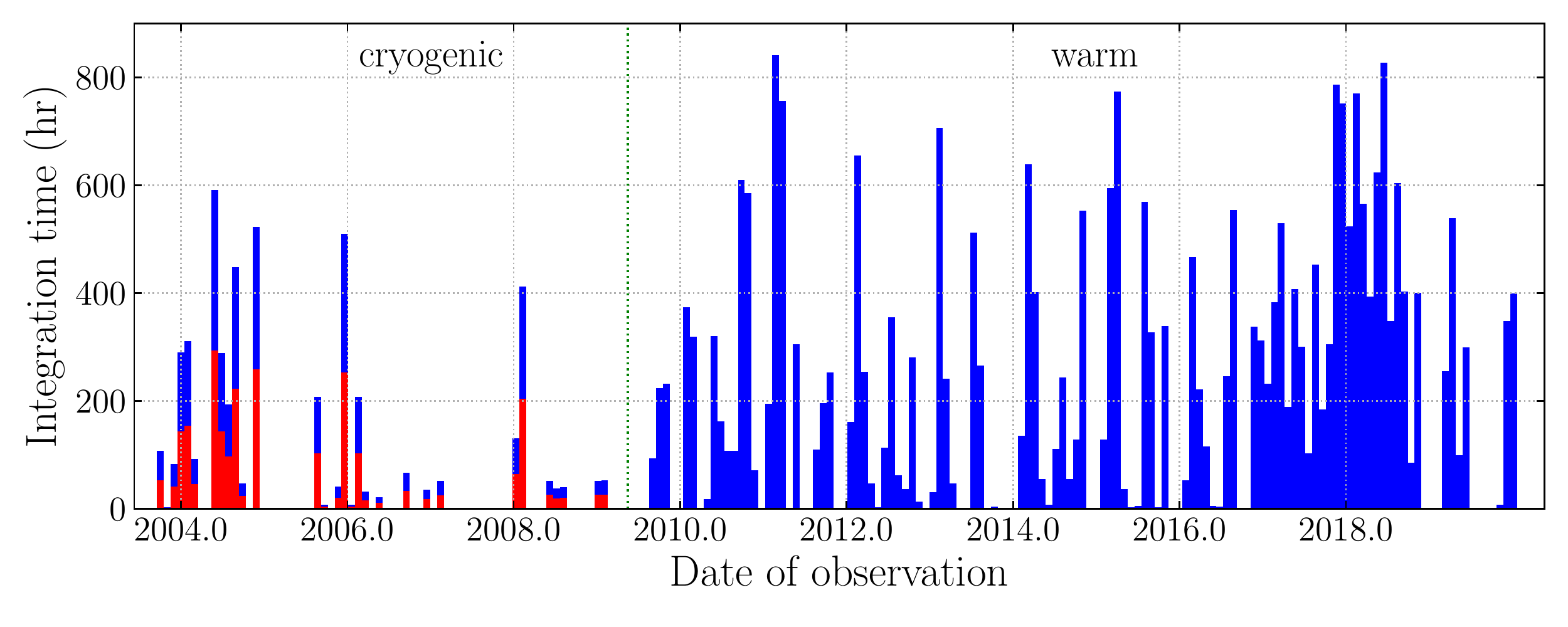}
    \caption{Histogram of the exposure time of the data analysed here (including the few discarded observations) using bins of 30 days. Our dedicated observations began in November 2016 and comprise most of the data after that date. The red part of each bar accounts for observations in channels 3 and 4, the blue part for those in channels 1 and 2; the vertical dotted line at 2009.37 indicates the end of the cryogenic mission. No observations are made in channel 3 and 4 after the end of the cryogenic mission.}
    \label{fig:obsTimeline}
\end{figure*}

All observations described here were made with IRAC. In brief, IRAC is a four-channel array camera on the \SpitzerST, observing simultaneously four fields slightly separated on the sky at $3.6$, $4.5$, $5.8$, and $8.0\mum$, known as channels 1--4, respectively. \Spitzer science observations began in August 2003 but observations in channels 3 and 4 ceased once the on-board cryogen was exhausted (May 15, 2009). During the following `warm mission' phase, channels 1 and 2 continued to operate until the end of operations in late January 2020, albeit with somewhat lower but still comparable performance. The earliest observations presented here are archival observations that were obtained in September 2003; the observations of the dedicated Capak program began in 2017 and the ones of the dedicated Scarlata program began in 2019.  The dedicated observations continued until January 2020, shortly before the shutdown of the satellite. Figure \ref{fig:obsTimeline} shows a histogram of the integration time accumulated in bins of 30 days over the observing period. These observations account for almost 1.5 million frames, a total integration time of \num{34\,000}\,hr, all channels combined, and a total on-target time, omitting overheads, of just over \num{15\,600}\,hr, or nearly 1.8\,yr, which is approximately 11\,\% of the \SpitzerST mission time. 

For our dedicated observations of EDF-N, EDF-S and EDF-F we adopted a consistent observing strategy that comprises blocks of $3\times3$ maps with a step size of 310\arcsec, a large three-point dither pattern and four repeats per position. Each block covers a $15\farcm1 \times 15\farcm1$ region with a coverage of $3\times4\times100\,$s exposures per pixel. The block centres are offset between passes in order to ensure uniform coverage and enable self-calibration. Each block forms an AOR, or Astronomical Observation Request, in IRAC jargon.  All other data included in our processing is archival data. It was obtained with a variety of observing strategies which we did not investigate in detail and which we do not attempt to summarise here.  In Appendix C we list the Program IDs of all the observations processed; in bold the ones of our dedicated observations. The combination of the archival data with our own dedicated data produces a spatially variable depth in most fields; this is discussed further in Sect.~\ref{sec:val}.

A total of 292 IRAC observing programs are used in this work. Table \ref{tab:Programs} lists the ten largest programs in terms in terms of observing time together with the PI of the program, the field concerned and program's total integration time.  

\begin{table*}[thb!] 
\label{tab:Programs} 
\centering \renewcommand{\arraystretch}{1.25} 
\caption{The ten largest programmes, by Program ID}
\begin{tabular}{ r l l r l} \hline \hline
 PID   & Principal Investigator  & Field &  Time (hr) &  Reference \\ \hline
61041  & Giovanni Fazio    & XMM   &   847 &  SEDS; \cite{ashby_seds_2013} \\
61040  & Giovanni Fazio    & HDFN  &   914 &  SEDS; \cite{ashby_seds_2013} \\
14235  & Claudia Scarlata  & EDF-S &  1086 &  this paper \\
  169  & Mark Dickinson    & HDFN  &  1104 &  GOODS; \cite{Labbe2015} \\
10042  & Peter Capak       & XMM   &  2033 &  this paper \\
90042  & Peter Capak       & COSM  &  2167 &  this paper \\
13094  & Ivo Labbe         & COSM  &  2483 &  GOODS; \cite{Labbe2015} \\
11016  & Karina Caputi     & COSM  &  3021 &  SMUVS; \cite{ashby_seds_2013} \\
13058  & Peter Capak       & EDF-F &  3162 &  this paper \\
13153  & Peter Capak       & EDF-N &  4625 &  this paper \\ \hline
\end{tabular} 
\end{table*}

All observations are summarised in Table \ref{tab:obs} which gives, for each field and channel, the number of frames (Data Collection Events or DCEs in IRAC terminology) used to produce the mosaics (note that this can be lower than the number of frames downloaded as some were discarded, see Sect.~\ref{sec:pro}) together with the total observing time. For channels 1 and 2, on the left side of the table, the information is subdivided into the cryogenic part and the warm part of the mission.


\begin{table*}[ht]
\caption{Valid observations}
\centering
\renewcommand{\arraystretch}{1.25}  
\setlength{\tabcolsep}{4pt}         
\begin{tabular}{l c r r r r r r r || c r r} 
\hline \hline
Field & Ch. & \multicolumn{2}{c}{cryo} & \multicolumn{2}{c}{warm} & \multicolumn{2}{c}{total} & & Ch. & \multicolumn{2}{c}{total} \\  \hline
 & & Num & Time & Num & Time & Num & Time & & & Num & Time \\ 
 \hline
EDF-N  &  1 &   5\,859 &   52 & 113\,521 & 2\,380 &  119\,380 &  2\,432 & & 3 &   5\,856 &   52 \\  
EDF-N  &  2 &   5\,857 &   52 & 113\,204 & 2\,467 &  119\,061 &  2\,519 & & 4 &   7\,667 &   50 \\ \hline 
EDF-F  &  1 &  14\,299 &  363 & 105\,781 & 2\,672 &  120\,080 &  3\,035 & & 3 &  14\,301 &  363 \\  
EDF-F  &  2 &  14\,299 &  363 & 105\,779 & 2\,764 &  120\,078 &  3\,127 & & 4 &  29\,686 &  352 \\  \hline
EDF-S  &  1 &   n/a    & n/a  &  21\,982 &    534 &   21\,982 &     534 & & 3 &      n/a &  n/a \\
EDF-S  &  2 &   n/a    & n/a  &  21\,982 &    552 &   21\,982 &     552 & & 4 &      n/a &  n/a \\ \hline
COSMOS &  1 &   7\,014 &  185 & 191\,072 & 4\,886 &  198\,086 &  5\,071 & & 3 &   7\,011 &  185 \\  
COSMOS &  2 &   7\,013 &  185 & 191\,031 & 5\,052 &  198\,044 &  5\,237 & & 4 &  13\,894 &  179 \\  \hline
EGS    &  1 &   4\,673 &  192 &  44\,101 &    551 &   48\,774 &     743 & & 3 &   4\,672 &  192 \\  
EGS    &  2 &   4\,673 &  192 &  44\,101 &    569 &   48\,774 &     761 & & 4 &  14\,535 &  186 \\ \hline 
HDFN   &  1 &   6\,253 &  298 &  36\,485 &    930 &   42\,738 &  1\,228 & & 3 &   6\,252 &  298 \\  
HDFN   &  2 &   6\,253 &  298 &  36\,485 &    962 &   42\,738 &  1\,260 & & 4 &  22\,496 &  288 \\ \hline
XMM    &  1 &  10\,264 &  154 &  98\,027 & 2\,410 &  108\,291 &  2\,564 & & 3 &  10\,265 &  154 \\  
XMM    &  2 &  10\,265 &  154 &  98\,030 & 2\,495 &  108\,295 &  2\,649 & & 4 &  14\,321 &  151 \\ 
\hline
\end{tabular}
\vspace*{0.1ex}
\tablefoot{
Here `Num' is the number of frames used, and `Time' is the integration time, in hours, they contribute. The left part of the table is for channels 1 and 2, split between cryogenic and warm mission, the right part is for channels 3 \& 4 which were used during the cryogenic mission only. Note that the EDF-S field was observed only during the warm mission.}
\label{tab:obs}
\end{table*}

\section{Processing} 
\label{sec:pro}

\subsection{Pre-processing and calibration}
\label{sec:preprocessing}

Processing begins with the \textit{Level 1} data products generated by the Spitzer Science Center via their `Basic Calibrated Data' pipeline \citep{10.1117/12.2233804}, which were downloaded from the NASA/IPAC Infrared Science Archive (IRSA\footnote{\url{https://irsa.ipac.caltech.edu}}). They have had all well-understood instrumental signatures removed, have been flux-calibrated in units of MJy\,sr$^{-1}$, and are delivered with an uncertainty image and a mask image; they are described in detail in the IRAC Instrument Handbook\footnote{\url{ https://irsa.ipac.caltech.edu/data/SPITZER/docs/irac/iracinstrumenthandbook/home}}. More precisely, we begin from the `corrected basic calibration data' products, which have file extensions \texttt{.cbcd} for the image, \texttt{.cbunc} for the uncertainty, and \texttt{.bimsk} for the mask. The files are grouped by AORs, namely sets of a few to several hundred DCEs obtained sequentially. All frames are \(256 \times 256\) pixels, the pixels are $1\farcs2$ wide, and the image file header contains the photometric solution and an initial astrometric solution. 

The processing is done region by region. A first pass over the files is used to check the headers for completeness and to discard a few incomplete AORs, which accounts for most of the differences in the number of frames listed in Table~\ref{tab:obs} between channels 1 and 2, or 3 and 4. This is followed by the correction of the `first frame' bias effect\footnote{\url{https://irsa.ipac.caltech.edu/data/SPITZER/docs/irac/iracinstrumenthandbook/26/}}. Next, the positions and magnitudes of WISE (\citep{wright_wide-field_2010}, \citep{mainzer_2011}) and \Gaia DR2 \citep{gaia_collaboration_gaia_2018} sources falling within the field are downloaded. The \Gaia sources are first `projected' to their location at the time of the observations using the \Gaia proper motions. Next they are identified on each IRAC frame, their observed fluxes and positions determined in each frame using the \texttt{APEX} software (the point-source extractor in \MOPEX\footnote{\url{https://irsa.ipac.caltech.edu/data/SPITZER/docs/dataanalysistools/tools/mopex/}}) in forced-photometry mode, and the positions are used to update the astrometric solution of each frame. There are typically 30--40 \Gaia DR2 sources available for each frame. In channels 1 and 2 most of them are detected and used for the astrometric correction. In the longer-wavelength channels 3 and 4 only a few sources in total are detected and usable but that is still sufficient to determine an astrometric solution with negligible distortion as shown in Sect.~\ref{sec:astro-pre}.  

An attempt was made to subtract bright stars in order to recover faint sources in their wings. For each AOR a model star built from the template PSFs described in the IRAC Instrument Handbook\footnote{\url{https://irsa.ipac.caltech.edu/data/SPITZER/docs/irac/iracinstrumenthandbook/19/}} (see Fig.~4.9 there) is scaled to the median of the fluxes of the star measured in that AOR, and is subtracted from each frame (of the AOR).  Different templates are available for each filter and separately for the cryogenic and the warm missions.  While this procedure worked quite well for moderately bright stars (which are of course the vast majority and which represent only a small loss in area), it introduced significant artefacts around the (few) very bright stars in the final mosaics. These artefacts included diffraction spikes corrected only out to a certain distance (out to where the template extends beyond the frame), other edge effects, and the subtraction of the core of bright galaxies. For these reasons the bright-star subtraction was not performed and the bright stars are left as they are.

\subsection{Stacking and image combination} \label{sec:combination}

In the next step we compute a median image for all frames within an AOR which corrects for persistence in the detectors and also for any residual first-frame pattern that introduces structure in the background. In parallel, a background map is also created by iteratively clipping objects and masking them, and finally that background is subtracted from each frame of the AOR. 

The final processing steps consist of resampling the background-subtracted frames onto a common grid with a scale of $0\farcs6\,\rm{pix}^{-1}$, i.e., half the instrument pixel size, that covers all data in all channels and which is the same in all channels. We experimented with two MOPEX interpolation schemes to produce our final mosaics. We first tried the ‘drizzling’ \citep{2002PASP..114..144F} scheme in which the final value of the output pixels is computed by considering the contribution of each input pixel in a smaller pixel grid in the output image. This procedure has excellent noise properties (it does not suffer from correlated pixels) when many input frames are available, but with few input frames it can produce artefacts in the output images. The second, simpler approach is to compute the value of each output pixel as a linear combination of the input pixel values. Although this procedure produces correlated noise, it works reliably for all the fields considered in this work which can have widely varying numbers of input images. Noise correlations can be estimated through simulations or by comparing sources in our drizzled and non-drizzled images. These comparisons show that the linear interpolation procedure leads to an underestimation of aperture magnitude errors by $30-40\%$ while the magnitudes themselves are unaffected.

Next, we use \MOPEX{} to produce an average-combined image while rejecting outliers and excluding masked regions. The stacking pipeline also produces the following ancillary characterisation maps: (1) an uncertainty map produced by stacking the input uncertainty maps using the same shifts as for the signal stack, (2) a coverage map giving the number of frames contributing to each pixel, and (3) an exposure time map giving the total exposure time per pixel. As the exposure times are not the same for all the observing programmes, these last two maps are not simply scaled versions of each other.

\subsection{On the spatial variation of the PSF in the stacks} \label{PSFshape}

The observations described here were made at many different satellite position angles (PAs), and thus when the images are stacked they must be rotated back to North upwards.  This has the effect of rotating the PSF, which is fixed in the satellite's reference frame.  Since the PSF is not rotationally symmetric, due in particular to the diffraction spikes, the stacked image of a star will depend on when it was observed.  As all parts of the stack were not observed at the same time (or at the same PA), the PSF varies spatially in the stack.


The COSMOS field, which is near the Equator, was observable only at specific times and therefore with a very restricted range of PAs; the PSF in the COSMOS stacks is thus quite homogeneous. But in the EDF-N, which was in a continuous viewing zone, observations were obtained at many different PAs, yielding more complicated and more spatially variable PSF.  This effect is very important for PSF-based photometry: the PSF at each position of the stack has to be reconstructed by stacking the nominal PSF at the PAs of the observations at that position, as did e.g., Labbe et al. (\citeyear{Labbe2015}) for the GOODS-South and HUDF fields, and also Weaver et al. (submitted) for the production of the COSMOS2020 catalogue.  The latter used the \texttt{PRFmap} code by Andreas Faisst, available at \url{https://github.com/cosmic-dawn/prfmap} for that purpose.  While doing such photometry is beyond the scope of this paper, we nevertheless provide, for each stack, a table of the PAs of each frame used in the stack. For completeness those tables also contain the frame coordinates, the MJD of the observation, and the exposure time; see Appendix~\ref{App:Prods} for more details.

\subsection{Products} \label{sec:products}

\begin{figure}[thb!]
    \includegraphics[width=0.95\hsize]{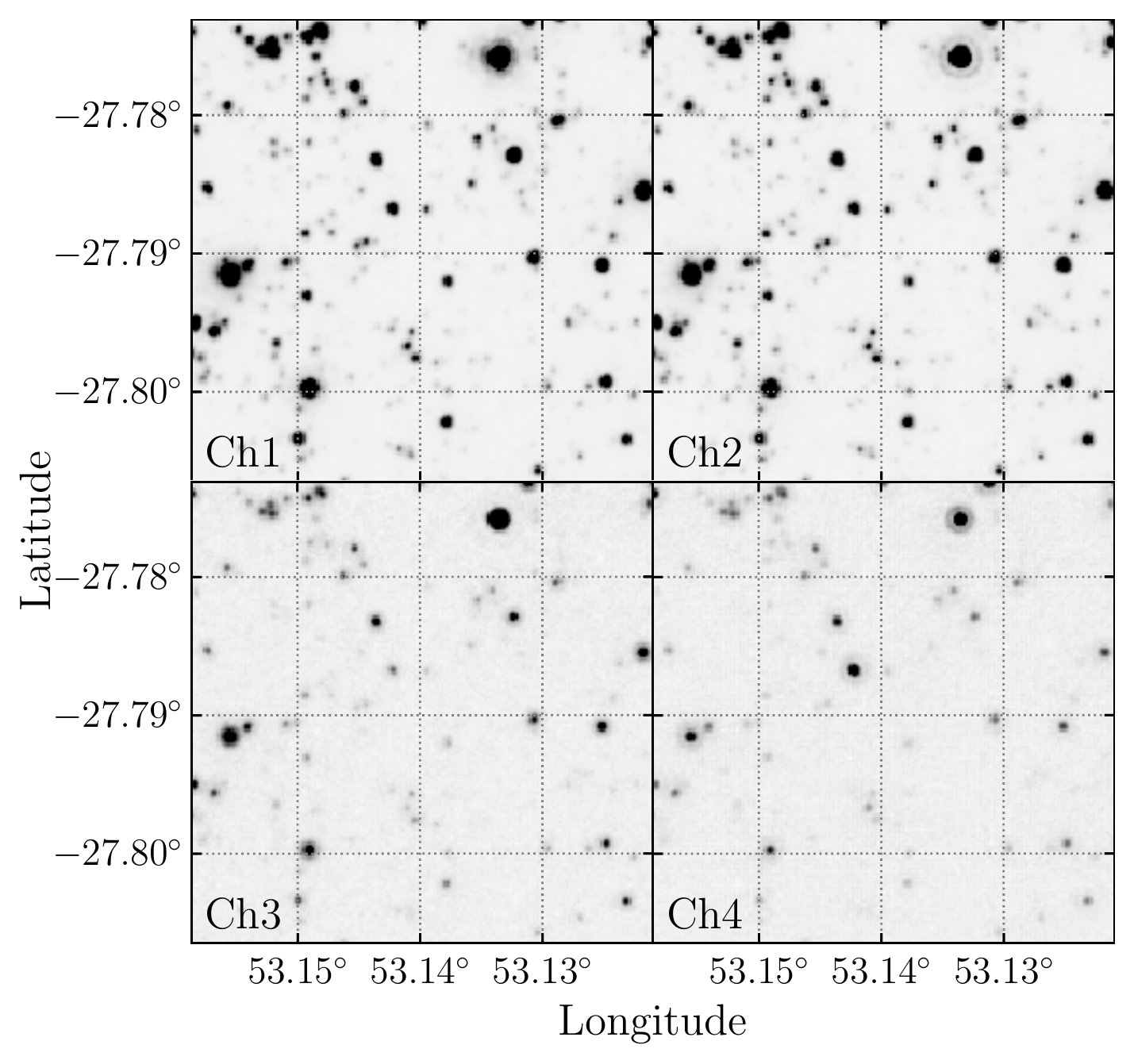}
    \caption{Detail of EDF-F mosaic in the region near that of maximum exposure time, which here is the same for all four channels. Images are $200\times 200$~pixels, or $2\arcmin \times 2\arcmin$. Display levels are $-\sigma$ to $+8\sigma$, where $\sigma$ is the standard deviation of the sky pixels, which is $\sim 0.005\,\rm{MJy}\,\rm{sr}^{-1}$ for channels 1 \& 2, and $\sim 0.013\,\rm{MJy}\,\rm{sr}^{-1}$ for channels 3 \& 4. }
    \label{fig:zoom1}
\end{figure}

As an example of the data quality, Fig.~\ref{fig:zoom1} shows a zoomed section of the EDF-F mosaic in the four channels near the region of maximum coverage.  We do not provide here figures of the full mosaics as they would be physically too small to show anything informative other than the overall coverage.

\begin{figure}[bht!]
    \includegraphics[width=0.95\hsize]{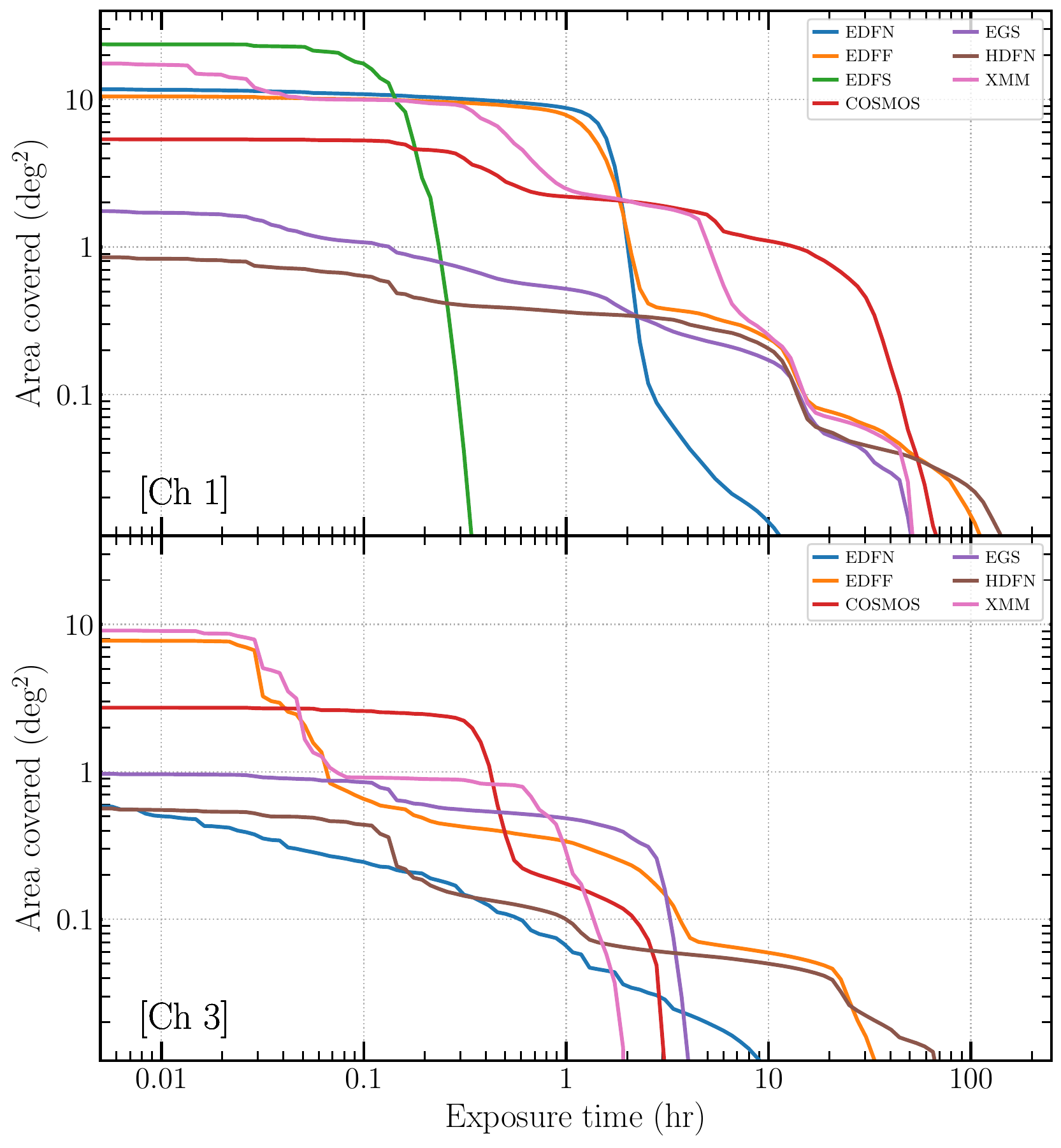}
    \caption{Cumulative area coverage as a function of exposure time for channels 1 and 3, for all fields. The figures for channels 2 and 4 are similar to the ones above, as explained in the text.}
    \label{fig:coverage}
\end{figure}

Maps of the integration time per pixel for channels 1 and 3 of all the fields are presented in Appendix~\ref{App:CovMaps}. Since channel 2 is observed together with channel 1, and similarly for channels 4 and 3, the paired channels have very similar coverage, albeit slightly shifted in position. The 10\,deg$^2$ circular area of EDF-N and EDF-F and the 20\,deg$^2$ pill-shaped area of EDF-S are easily seen on those figures. Also, and with the exception of EDF-S, for which there are only observations done specifically for this programme and no archival data, the integration time per pixel, and consequently the depth reached, is far from uniform, with only a small part of the total area of each field having been observed for more than a few hours. In fact, the median integration time per pixel is larger than 1\,hr for only two fields.  Table~\ref{tab:int-times} gives the median and maximum pixel integration time for each field and each channel.

\begin{table}[thb!] 
\label{tab:int-times} 
\centering \renewcommand{\arraystretch}{1.25}    
\setlength{\tabcolsep}{3pt}        
\caption{Median and maximum pixel integration time in hours}
\begin{tabular}{l r r r r r r r r} \hline \hline
Field  &  \multicolumn{2}{c}{ch1} & \multicolumn{2}{c}{ch2} &
 \multicolumn{2}{c}{ch3} & \multicolumn{2}{c}{ch4} \\  \hline
COSMOS  &  0.51 &  93.7  & 0.50 &  97.1 & 0.38 &  5.1  & 0.38 &  5.5 \\
EDF-F   &  1.33 & 199.7  & 1.33 & 149.5 & 0.03 & 47.3  & 0.03 & 54.2 \\
EDF-N   &  1.47 &  23.4  & 1.56 &  21.3 & 0.04 & 20.4  & 0.04 & 19.6 \\
EDF-S   &  0.13 &   0.5  & 0.16 &   0.5 &  --  &  --   &  --  &  --	 \\
EGS     &  0.16 &  71.1  & 0.16 &  71.6 & 0.93 &  5.4  & 0.95 &  4.8 \\
HDFN    &  0.16 & 236.2  & 0.16 & 224.4 & 0.13 & 91.2  & 0.13 & 95.2 \\
XMM     &  0.31 &  65.9  & 0.33 &  67.1 & 0.04 &  2.0  & 0.04 &  2.0 \\
\hline \end{tabular} 
\end{table}

That variation of area covered as a function of exposure time for channels 1 and 3 and for all fields is shown graphically in Fig.~\ref{fig:coverage} which presents a cumulative histogram of the area covered vs. exposure time. The intersection of the curve with the vertical axis thus gives the total area covered for that field and these areas are also listed in Table~\ref{tab:nareas}. EDF-S is the most uniformly observed field and it covers the largest area, but it is also the shallowest, with only 0.1 hr per pixel on average, and it is also the only field with no channel 3 and 4 data.  EDF-F and EDF-N reach the target coverage of 10\,deg$^2$ with about 1\,hr of exposure time, with the latter showing deeper coverage over smaller zones.  The other fields were covered by many observing programs with different objectives and which covered specific areas to different depths. The combination of these programs with our own yields a curve with many plateaus. Finally, there are a few small parts of the EDF-F and HDFN fields that have more than 100\,hr of exposure time. 

\begin{table}[thb!] 
\label{tab:nareas} 
\centering \renewcommand{\arraystretch}{1.25} 
\setlength{\tabcolsep}{4pt}        
\caption{Location and area, in deg$^2$, covered in each field}
\begin{tabular}{l r r r r r r r} \hline \hline
Field   &   RA\   & Dec\  &  & ch1\  & ch2\  & ch3\  & ch4\  \\ \hline
EDF-N   & \ra{17;58;} & \ang{ 66;36;} & & 11.74  & 11.54  & 0.61 & 0.62 \\
EDF-F   & \ra{ 3;32;} & \ang{-28;12;} & & 10.52  & 11.05  & 7.78 & 7.77 \\
EDF-S   & \ra{ 4;05;} & \ang{-48;30;} & & 23.60  & 23.14  &  --  &  --  \\
COSMOS  & \ra{10;00;} & \ang{  2;12;} & &  5.37  &  5.46  & 2.72 & 2.72 \\
EGS     & \ra{14;19;} & \ang{ 52;42;} & &  1.76  &  1.80  & 0.97 & 0.98 \\
HDFN    & \ra{12;37;} & \ang{ 62;24;} & &  0.91  &  0.91  & 0.57 & 0.63 \\
XMM     & \ra{ 2;27;} & \ang{ -4;36;} & & 17.54  & 17.48  & 9.09 & 9.10 \\
\hline \end{tabular} 
\end{table}

\subsection{Final sensitivities} \label{sec:sensitivities}

\begin{figure}[tbh]
    \centering
    \includegraphics[width=0.95\hsize]{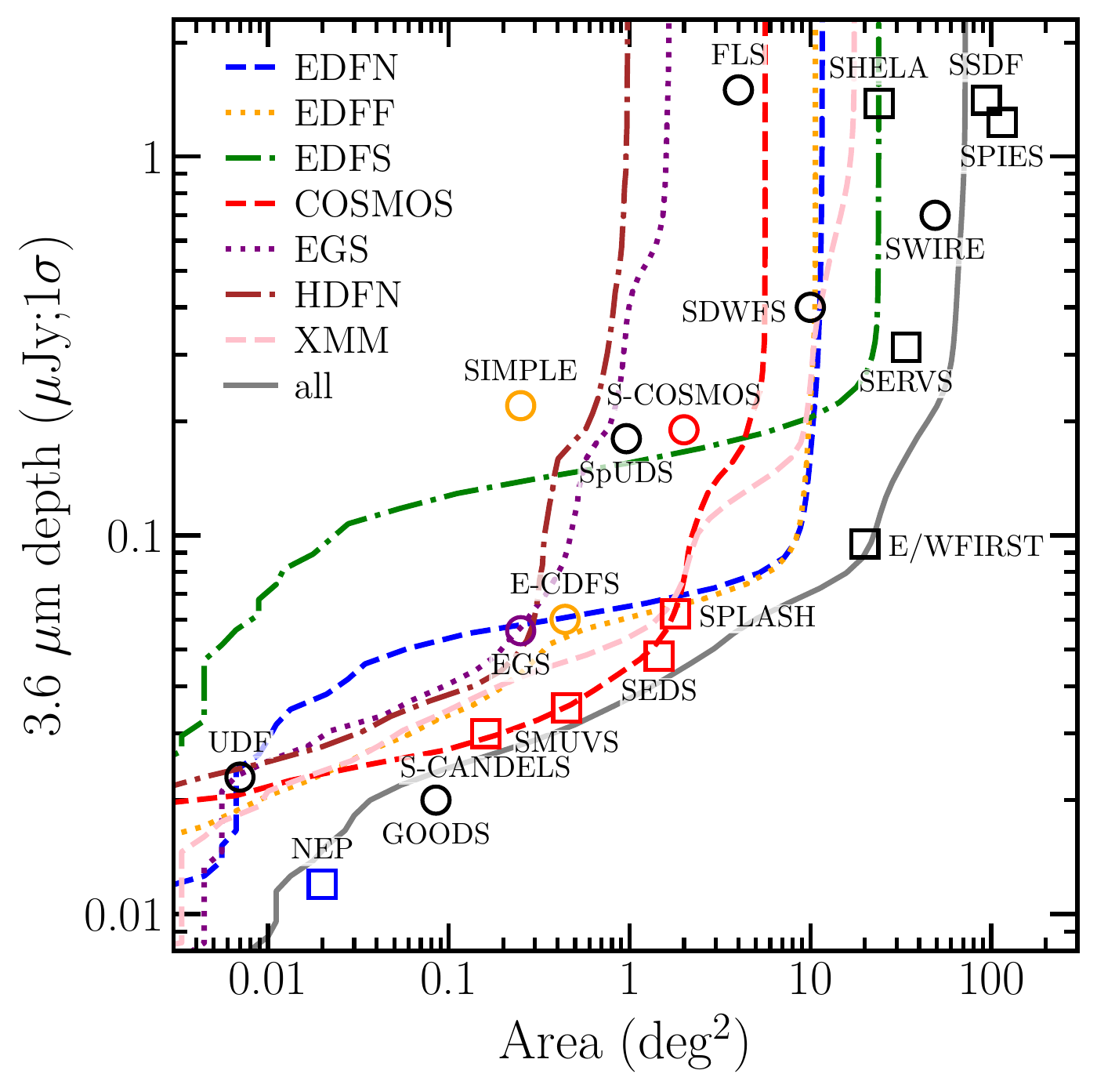}
    \caption{Sensitivity of the \textit{Spitzer}/IRAC channel 1 data as a function of cumulative area coverage. The coloured lines illustrate $1\,\sigma$ depths measured in empty $2\farcs5$ diameter apertures in each field. The grey solid line is the total area observed to a given depth summed over different surveys. The data points indicate point-source sensitivities at $1\sigma$ compiled in \cite{ashby_spitzer_2018} (note that some of these data are included in our stacks). The circles and squares represent surveys executed during cryogenic and warm missions, respectively.}
    \label{fig:depth_area}  
\end{figure}

We estimate the sensitivities of the stacked images by measuring the flux in circular $2\farcs5$ diameter apertures randomly placed across each image after masking the regions with detected objects using the \texttt{SExtractor} \citep{bertin_sextractor_1996} segmentation map. The sensitivity is then computed as the standard deviation of these fluxes (3$\sigma$ clipped). This procedure is done in $200\times200$ pixel cells (4\,arcmin$^2$).  Figure \ref{fig:depth_area} shows the cumulative area covered as a function of sensitivity for the channel 1 mosaics. Note the similarity between this figure and the top panel of Fig.~\ref{fig:coverage} once the latter is rotated by 90 degrees.  The solid line shows our total depth, summed over all our survey fields. Also shown in the figure are the published sensitivities of the surveys that are included in our data and analyses. Generally, our measured sensitivities are consistent with literature measurements for surveys of equivalent exposure time.

\section{Validation and quality control} \label{sec:val}

As part of our validation process we compare photometry and astrometry of sources in our stacks with reference catalogues and also extract number counts that can be compared to previous works.

\subsection{Catalogue extraction} \label{sec:cat_ext}

We begin by extracting source catalogues from the channels 1 and 2 stacks of all fields using \texttt{SExtractor}. We adopt the usual approach of searching for objects that contain a minimum number of connected pixels above a specified noise threshold (in this case $2\,\sigma$) and measuring their aperture magnitudes. In the case of our moderately deep IRAC data, where many sources are blended due to the large IRAC PSF, this approach is known to miss faint sources. However, these faint sources are not required for our quality assessment purposes and a shallower catalogue is entirely sufficient. 
\texttt{SExtractor} estimates a global background on a grid with mesh size of $32\times32$ pixels (recall that pixels are $0\farcs6$ wide). This background is smoothed with a $5\times5$ pixel Gaussian kernel with ${\rm FWHM}=\ang{;;1.5}$. For each source, the flux is measured within a circular aperture of $\ang{;;7}$ diameter and a local background is estimated within an annulus of width $32$~pixels around the isophotal limits. The measured fluxes were converted from MJy/sr to AB magnitude using a zero-point of $21.58$ (which accounts for a zero-magnitude flux of 3631\,Jy and a pixel size of $0\farcs6$\footnote{\url{https://irsa.ipac.caltech.edu/data/SPITZER/docs/irac/iracinstrumenthandbook/19/}}), and the latter were converted to total magnitude using the aperture corrections given in the IRAC Instrument Handbook for the warm mission ($-0.1164$ and $-0.1158$ for channel 1 and channel 2 respectively), which covers the vast majority of the data, while the correction for the cryogenic mission differs at only the 1--2\% level\footnote{\url{https://irsa.ipac.caltech.edu/data/SPITZER/docs/irac/calibrationfiles/ap_corr_warm/}}.  A list of relevant \texttt{SExtractor} parameters used for the catalogue extraction can be found in Table~\ref{tab:SEparams}.

\begin{table}[hbt!] 
\caption{\texttt{SExtractor} parameters used for detection and photometry.}
\label{tab:SEparams}
\begin{center}
\renewcommand{\arraystretch}{1.00} 
\setlength{\tabcolsep}{4pt} 
\begin{tabular}{l c} \hline \hline
Parameter name & Value \\ \hline
\texttt{DETECT\_MINAREA} & 5  \\ 
\texttt{DETECT\_MAXAREA} & 1000000 \\ 
\texttt{THRESH\_TYPE} & \texttt{RELATIVE} \\
\texttt{DETECT\_THRESH} & 2 \\
\texttt{ANALYSIS\_THRESH} & 2 \\ 
\texttt{FILTER\_NAME} & \texttt{gauss\_2.5\_5x5.conv} \\ 
\texttt{DEBLEND\_NTHRESH} & 32 \\ 
\texttt{DEBLEND\_MINCONT} & 0.00001 \\ 
\texttt{BACK\_SIZE} & 32  \\
\texttt{BACKPHOTO\_THICK} & 32 \\
\texttt{BACK\_FILTERSIZE} & 3 \\
\texttt{BACKPHOTO\_TYPE}  & \texttt{LOCAL} \\
\texttt{MAG\_ZEROPOINT} & 21.58 \\
\texttt{PHOT\_AUTOPARAMS} & 2.5,5.0 \\
\texttt{PIXEL\_SCALE} & 0.60 \\
\hline 
\end{tabular}\end{center}\end{table}

\subsection{Astrometric and photometric validation} \label{sec:astro-pre}

Using the catalogues extracted above, we evaluate the astrometric accuracy of our stacked images. For each field we cross-match sources with magnitude $16 <\rm{[3.6]} < 19$ within $1\arcsecond$ of their counterparts in the \Gaia DR2 catalogue. This magnitude range was adopted to ensure only bright, non-blended sources were chosen. We now present a detailed analysis for EDF-N but other fields are similar. 

\begin{figure}[hbt!]
    \includegraphics[width=0.95\hsize]{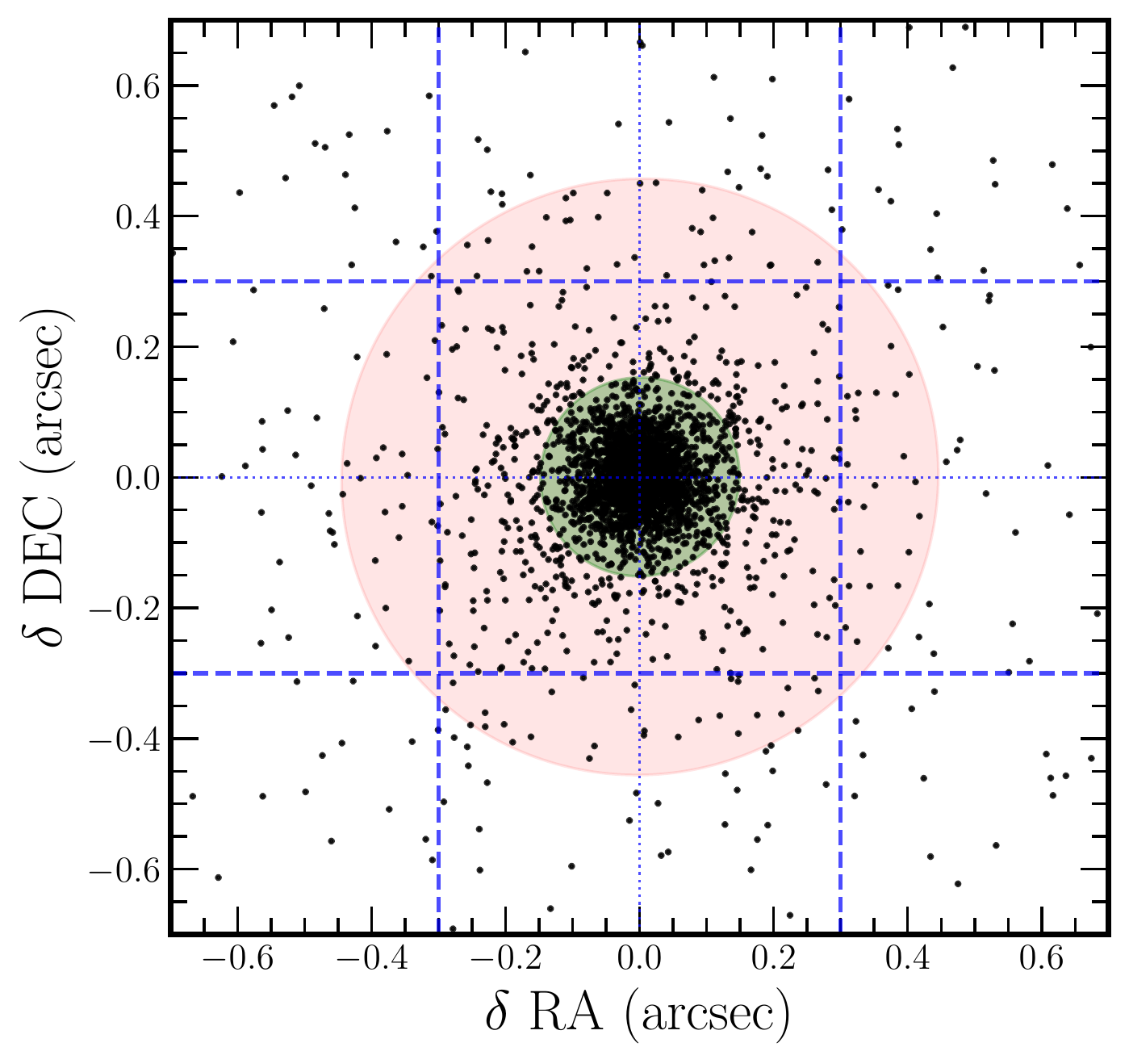}
    \caption{The difference between the reference and the measured position, in arcseconds, of \Gaia DR2 catalogue sources with $16<\rm{[3.6]}<19$ total magnitudes extracted from the EDF-N channel 1 mosaic. The blue dashed lines indicate the size of one mosaic pixel. The blue dotted lines go through the origin. The shaded regions are ellipses containing $68\,\%$ and $99\,\%$ of all sources respectively. For clarity, only one in ten sources is plotted.} 
    \label{fig:2d-resid}
\end{figure}

\begin{figure}[hbt!]
    \includegraphics[width=0.95\hsize]{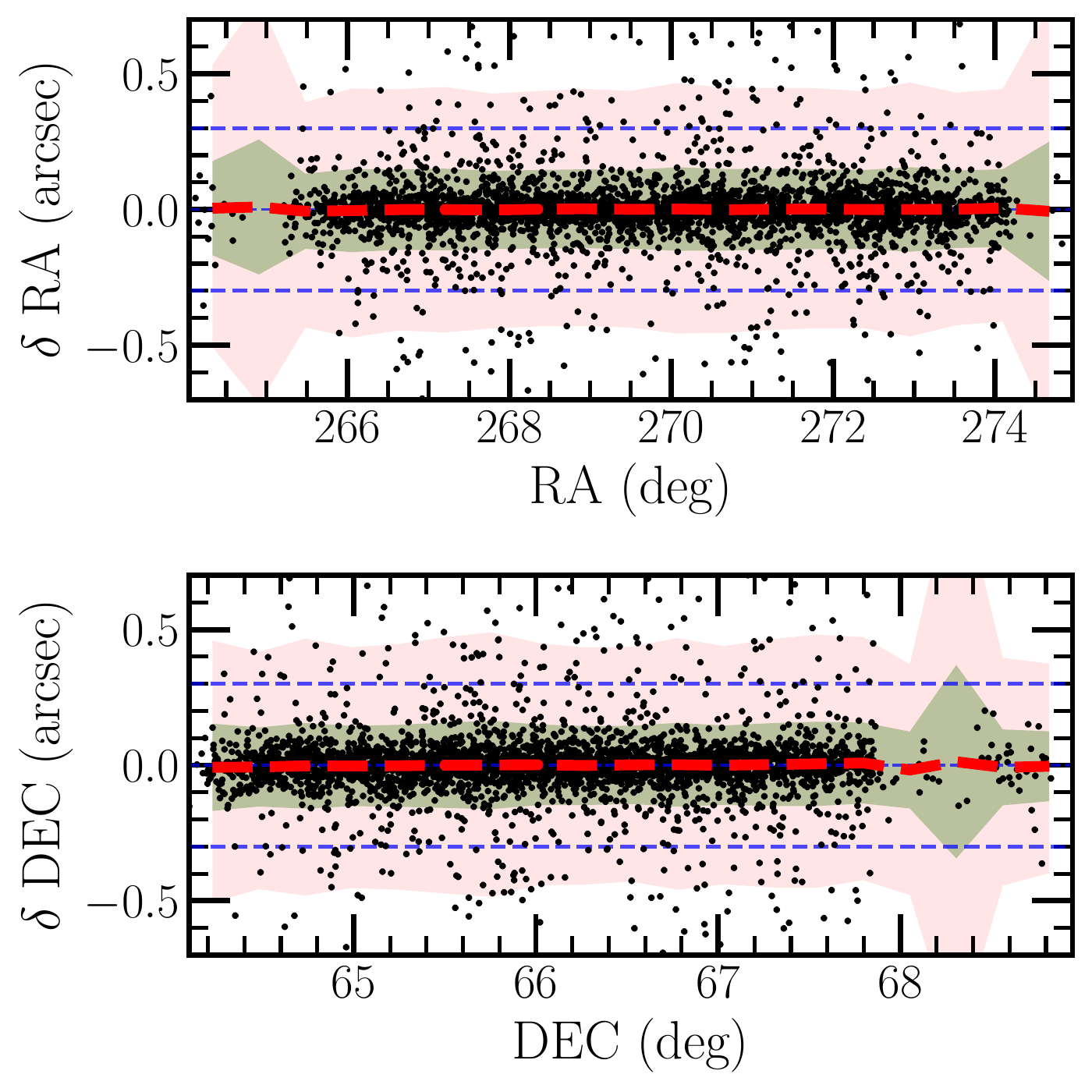}
    \caption{The difference between the reference \Gaia DR2 catalogue and the measured RA (top panel) and DEC (bottom panel) of sources in the EDF-N channel 1 mosaic with $16<\rm{[3.6]}<19$ total magnitudes as a function of the coordinate. The solid red line shows a running median computed in bins of 20 points, and the shaded areas indicate the regions containing $68\,\%$ and $99\,\%$ of all sources respectively. }
    \label{fig:1d-resid}
\end{figure}

Figure \ref{fig:2d-resid} shows the difference between reference and measured coordinates (for clarity, only one point in ten is shown). The heavy blue dashed line gives the size of one pixel in the stacked image (which is half the size of the instrument pixel). Similarly (again showing only one in ten points), Fig.~\ref{fig:1d-resid} shows, for each coordinate, the difference between the reference and the measured value as a function of position along the other coordinate. The thick red dashed line shows a running median computed over a bin containing 20 points. The flatness of this line indicates that there is no significant spatial variation in astrometric precision. Considering all fields, we find that the 1\,$\sigma$ precision (measured as the RMS of the difference between positions in our catalogue and those in \Gaia DR2) is $0\farcs15$. Furthermore, the median value is always $\lob 0\farcs1$, with the exception of the sparsely-covered HDF-N field where it is $\lob 0\farcs2$.

These measurements demonstrate that the astrometric solutions have been correctly applied to the individual images and that the combined images are free of residuals on a scale much smaller than an individual mosaic pixel, which is more than sufficient to measure precise infrared and optical-infrared colours. 

\begin{figure}[hbt!]
    \includegraphics[width=0.95\hsize]{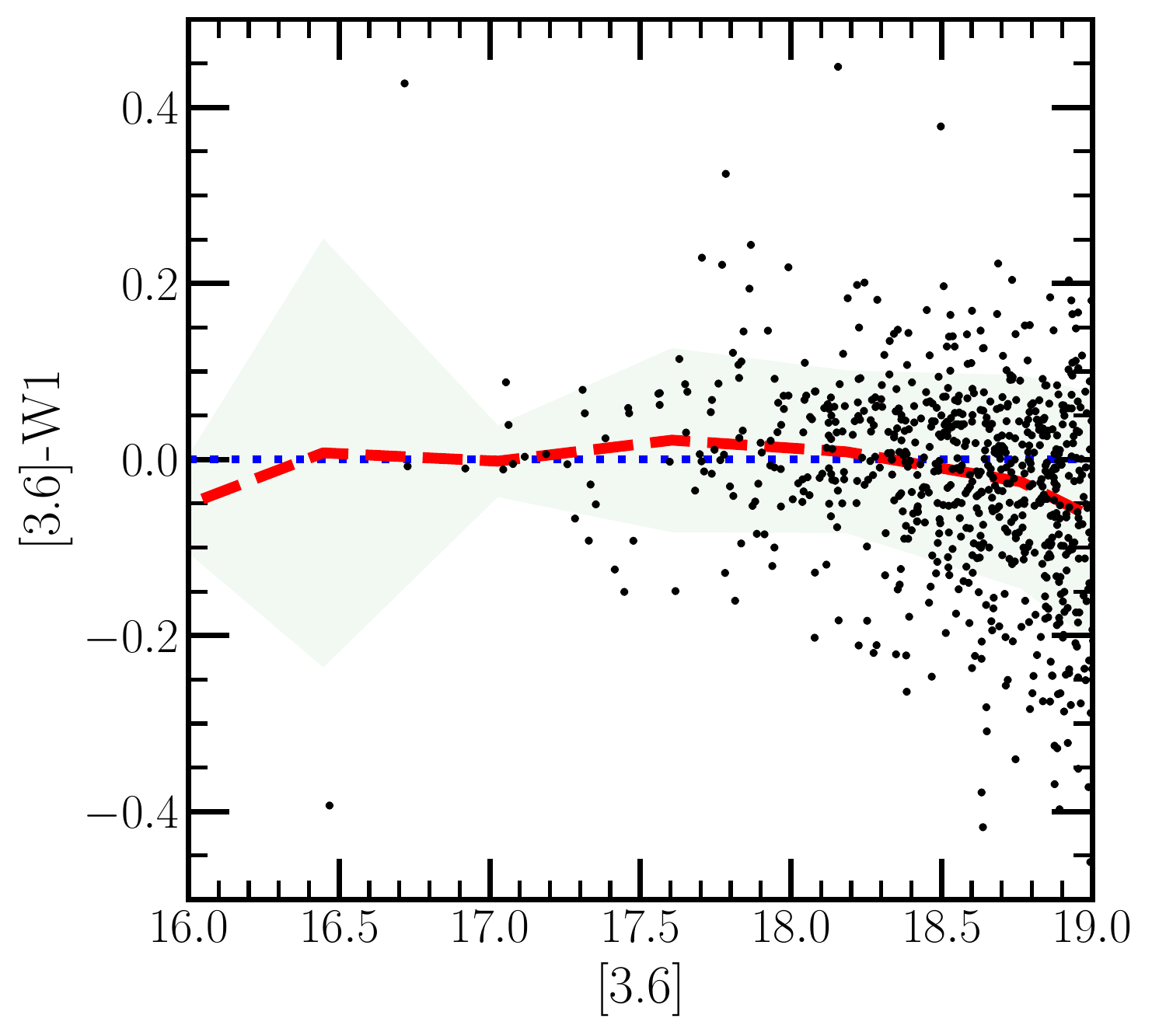}
    \caption{Photometric comparison with the WISE survey. The magnitude measured in $7\arcsec$ apertures for flat-spectrum objects ($[3.6]-[4.5] \sim 0$) is compared with W1 magnitudes in the ALLWISE survey. The shaded area represents the 68\% confidence interval.}
    \label{fig:magWiseComp}
\end{figure}

Finally, we perform a simple check on the photometric calibration of our mosaics. As described previously, individual images are photometrically calibrated by the Spitzer Science Center (SSC). Following the validation procedures outlined by the SSC, we compare magnitudes of objects in our catalogues with those in the WISE survey. Because of the difference between the WISE W1 and IRAC channel 1 filter profiles, we select objects with $[3.6]-[4.5]\sim0$. Figure \ref{fig:magWiseComp} shows the magnitude difference for the EDF-N field, and the agreement is excellent. Further comparisons with photometric measurements in previous COSMOS IRAC surveys can be found in the Appendix of Weaver et al. 

\subsection{Magnitude number counts} \label{sec:number-counts}

\begin{figure}[hbt!]
    \includegraphics[width=0.95\columnwidth]{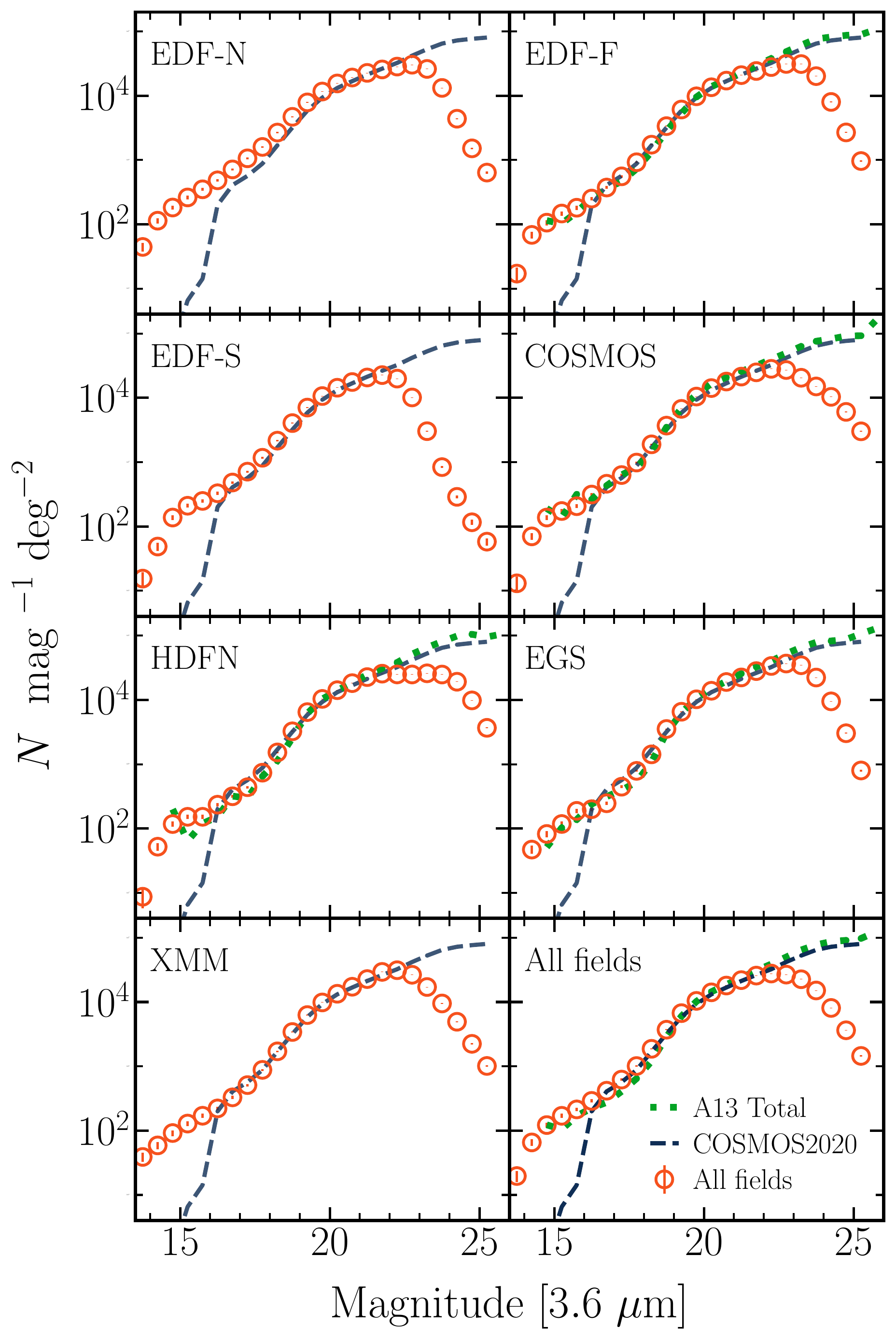}
    \caption{Magnitude number counts in channel 1 (red circles) together with COSMOS2020 (long dashed lines) and  \citet[hereafter A13; short dotted green lines]{ashby_seds_2013}. The bottom right panel shows the mean of all fields compared to , and the legend there applies to all panels.}
    \label{fig:NumCounts}
\end{figure}

 We compute the differential number counts in channel 1 in each field using the corrected $\ang{;;7}$ aperture magnitudes. Since the  IRAC PSF is too large to perform morphological source classification, we simply include all objects detected.  These are shown in Fig.~\ref{fig:NumCounts}, where the red circles with uncertainties present our measurements and the lines show the number counts from the literature; the bottom-right panel shows the mean of all fields. We compare our number counts with those presented in \citet{ashby_seds_2013} who also surveyed many of our fields and also with those computed using the new COSMOS2020 photometric catalogue (Weaver et al., submitted) which we use as a reference. 

There is a general agreement in the number counts in all the fields with \citet{ashby_seds_2013} and COSMOS2020 for $16 < [3.6] < 22$. At brighter magnitudes the COSMOS2020 counts drop off as bright sources were not included. At fainter magnitudes, our aperture-based catalogues are confusion-limited and thus incomplete. Conversely, the COSMOS2020 catalogue, which uses a high-resolution prior for the detection and a profile-fitting method for the measurement, is complete up to significantly fainter magnitudes.

\begin{figure}[hbt!]
    \centering
    \includegraphics[width=0.95\columnwidth]{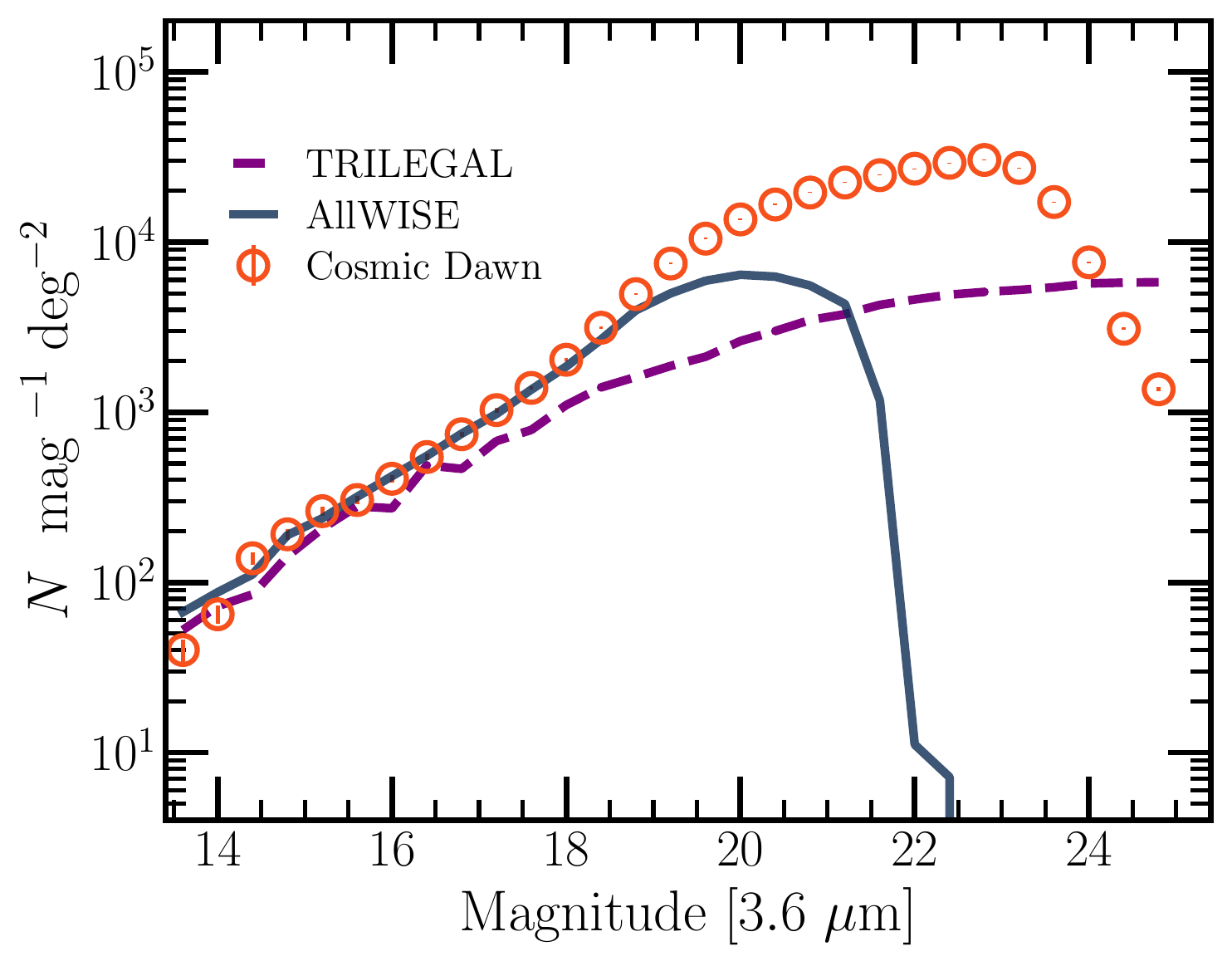}
    \caption{Magnitude number counts in the EDF-N field compared to AllWISE and the predicted stellar number counts from \texttt{TRILEGAL}.}
    \label{fig:NumCounts-NEP-TRILEGAL}
\end{figure}

EDF-N counts are slightly higher than the other fields at bright magnitudes. To investigate this difference we simulated a stellar catalogue of $1\,{\rm deg}^2$ centred on EDF-N using \texttt{TRILEGAL} \citep{Girardi2005_trilegal} and compared counts from this simulated catalogue with our observations, shown in Fig.~\ref{fig:NumCounts-NEP-TRILEGAL}. At bright magnitudes, where stars are expected to outnumber galaxies, our counts are in reasonable agreement with \texttt{TRILEGAL} predictions, and in excellent agreement with the number counts extracted from the AllWISE \citep{wright_wide-field_2010} catalogue for this field. These comparisons indicate that the difference between EDF-N and other fields is largely due to the higher density of stellar sources in there, consistent with its lower Galactic latitude.  

\section{Summary} 
\label{sec:conc}
We have presented the \Spitzer/IRAC mid-infrared component of the Cosmic Dawn Survey: an effort to complement the \Euclid mission's observations of deep and calibration fields with deep longer-wavelength data to enable high redshift legacy science. 

The survey consists of two major new programs covering the three \Euclid deep fields (EDF-N, EDF-F and EDF-S) and a homogeneous reprocessing of all existing data in \Euclid's four calibration fields (COSMOS, XMM, EGS and HDFN). We have processed new data together with all relevant archival data to produce mosaics of these fields covering a total of $\sim 71$\,deg$^2$ in IRAC channels 1 and 2.  Furthermore, the new mosaics are tied to the \Gaia astrometric reference system. The MIR data will be essential for a wide range of legacy science with \Euclid, including improved star/galaxy separation, more accurate photometric redshifts, determination of stellar masses of galaxies, and the construction of complete  galaxy samples at $z>2$ with well understood selection effects.   

We validated our final products by comparing catalogues extracted from channels 1 and 2 to external catalogues. In all fields, comparing with Gaia DR2, the residual astrometric uncertainty for sources with total magnitudes $16 <\rm{[3.6]} < 19$ is around $0\farcs15$ ($1\sigma$). Our photometric measurements are in excellent agreement with WISE photometry and our number counts are consistent with previous determinations.  

The Cosmic Dawn Survey \Spitzer survey presented here represents the first essential step in assembling the required multi-wavelength coverage in the \Euclid deep fields which are set to become some of the most important fields in extragalactic astronomy for the coming decade. Since the \Spitzer mission has finished, and all available data in these fields have been processed with the latest reduction pipeline, the resulting mosaics will remain the deepest and widest MIR imaging survey for the foreseeable future. No existing or approved future observatories are capable of obtaining such data. While JWST is more sensitive and has higher spatial resolution at these wavelengths, its mapping speed is too slow to cover comparable degree-scale areas.  

In the context of the Cosmic Dawn Survey, several programs are currently underway to add data at other wavelengths to the \Euclid deep fields and calibration fields. In particular deep optical data in the EDF-N and EDF-F are currently being obtained with the Subaru's Hyper-Suprime-Cam instrument as part of the Hawaii-Two-0 program (McPartland et al., in prep). These fields are also being targeted with high spatial resolution millimeter observations as part of the planned Large-scale Structure Survey with the Toltech Camera\footnote{ http://toltec.astro.umass.edu/} on the Large Millimeter Telescope \citep[LMT][]{Pope2019}. A deep $U$-band survey is also underway with the CFHT (Zalesky et al., in prep). EDF-S is being covered with K-band observations from the VISTA telescope (Nonino et al., private communication), and planning is ongoing to obtain optical data with the Vera C.~Rubin Observatory.  

The Cosmic Dawn Survey \Spitzer mosaics and associated products described here can be downloaded from the IRSA web site, Appendix~\ref{App:Prods} gives the details of the download site and the naming convention used. The community is encouraged to make use of them for their science.

\begin{acknowledgements}
We thank the \MOPEX{} support team for fixing issues that appeared when combining large numbers of files. This publication is based on observations made with the \SpitzerST, which is operated by the Jet Propulsion Laboratory, California Institute of Technology under a contract with NASA, and has made use of the NASA/IPAC Infrared Science Archive, which is funded by the National Aeronautics and Space Administration and operated by the California Institute of Technology.
This publication has also made use of data from the European Space Agency (ESA) mission \Gaia (\url{https://www.cosmos.esa.int/gaia}), processed by the \Gaia Data Processing and Analysis Consortium (DPAC,
\url{https://www.cosmos.esa.int/web/gaia/dpac/consortium}). Funding for the Gaia Data Processing and Analysis Consortium (DPAC) has been provided by national institutions, in particular the institutions participating in the \Gaia Multilateral Agreement. This publication makes use of data products from the Wide-field Infrared Survey Explorer (WISE), which is a joint project of the University of California, Los Angeles, and the Jet Propulsion Laboratory/California Institute of Technology, funded by the National Aeronautics and Space Administration. H.J.McC.~acknowledges support from the PNCG. This work used the CANDIDE computer system at the IAP supported by grants from the PNCG and the DIM-ACAV and maintained by S.~Rouberol. S.T.~and J.W.~acknowledge support from the European Research Council (ERC) Consolidator Grant funding scheme (project ConTExt, grant No.~648179). I.D.~has received funding from the European Union's Horizon 2020 research and innovation programme under the Marie Sk\l{}odowska-Curie grant agreement No.\ 896225. The Cosmic Dawn Center is funded by the Danish National Research Foundation under grant No. 140. H.Hildebrandt is supported by a Heisenberg grant of the Deutsche Forschungsgemeinschaft (Hi 1495/5-1) as well as an ERC Consolidator Grant (No. 770935).\AckEC

\end{acknowledgements}

\bibliographystyle{aa}    
\bibliography{./references_2.bib}

\begin{appendix}

\section{Delivered data products} \label{App:Prods}

The new mosaics and associated products can be obtained from the IRSA website at \url{https://irsa.ipac.caltech.edu/data/SPITZER/Cosmic_Dawn} (ATTN: the products will become available once the paper is accepted; the URL may be updated at publication time). The file naming convention for the stacks is as follows:
\begin{verbatim}
    CDS_{field}_ch{N}_{type}_v24.fits
\end{verbatim}
where \verb|field| is the field name, \verb|N| is the channel number, and \verb|type| is one of \begin{description} 
    \item \verb|ima|: for the flux image, 
    \item \verb|cov|: for the coverage in terms of number of frames used to build each pixel of the mosaic, 
    \item \verb|tim|: for the exposure time in sec of the pixel, and 
    \item \verb|unc|: for the uncertainty as determined from the standard deviation of the image pixels that contributed to the mosaic pixel.  \end{description}

Also, Table \ref{tab:details} gives the precise J2000 coordinates of the field tangent point in decimal degrees, the reference pixel corresponding to that tangent point, and size, in pixels, of the mosaics.  These values are the same for all channels of a field and for all the ancillary images.  The pixel scale is $0\farcs60$ per pixel for all mosaics.

\begin{table*}[tbh!] \label{tab:details}
\centering 
\renewcommand{\arraystretch}{1.25}  
\caption{Data products information}
\begin{tabular}{l r r r r r r} \hline \hline
Field & Longitude & Latitude & $x$-size & $y$-size & $x$-ref.pix & $y$-ref.pix \\  \hline
EDF-N    & 269.485804 &   66.590708 & 27\,410 & 30\,148 & 13\,705.55 & 15\,074.53 \\ 
EDF-F    &  53.062008 & --28.205431 & 23\,751 & 26\,204 & 11\,876.02 & 13\,102.29 \\ 
EDF-S    &  61.301724 & --48.496065 & 41\,676 & 33\,976 & 20\,838.59 & 16\,988.50 \\ 
COSMOS   & 150.178292 &    2.220994 & 15\,440 & 17\,804 &  7\,720.46 &  8\,902.40 \\ 
EGS      & 214.781187 &   52.720882 & 11\,278 & 13\,649 &  5\,639.32 &  6\,824.97 \\ 
HDFN     & 189.405434 &   62.373754 & 11\,813 &  1\,979 &  5\,907.03 &  8\,489.78 \\ 
XMM      &  34.101249 &  --4.598575 & 47\,583 & 25\,022 & 23\,791.97 & 12\,511.69 \\ 
\hline \end{tabular}
\tablefoot{Longitude and latitude are Equatorial and J2000, for the image tangent point. These values are valid for all four channels of each field and for their ancillary images.}
\end{table*}

The tables with the observation date, coordinates, position angles, and exposure times of the input frames are provided in IPAC format and are gzipped to reduce their size.  Their names are as follows:
\begin{verbatim}
    CDS_{field}_ch{N}_info_v24.tbl.gz
\end{verbatim}
The first few lines of the table for channel 1 of the EGS field are as follows:
\begin{verbatim}
  |          MJD|              RA|             DEC|               PA|ExpTime|
  |       double|          double|          double|           double| double|
  |          day|             deg|             deg|              deg|    sec|
  |         null|            null|            null|             null|   null|
   53822.6296863 214.458364468008 51.9912156620541 -126.246899270774     0.4 
   53822.6297117 214.458364468008 51.9912156620541 -126.247055616923    10.4 
   53822.6298641 214.458364468008 51.9912156620541 -126.246928305428    96.8 
   53822.6312156 214.383044470269 52.0556500561605 -126.304483006668    96.8 
\end{verbatim}
The coordinates are in degrees of longitude and latitude (Equatorial, J2000) and the PAs are measured eastward of North.

\section{Coverage maps} \label{App:CovMaps}

\begin{figure*}
    \centering
    \includegraphics[width=24.5cm, angle=90]{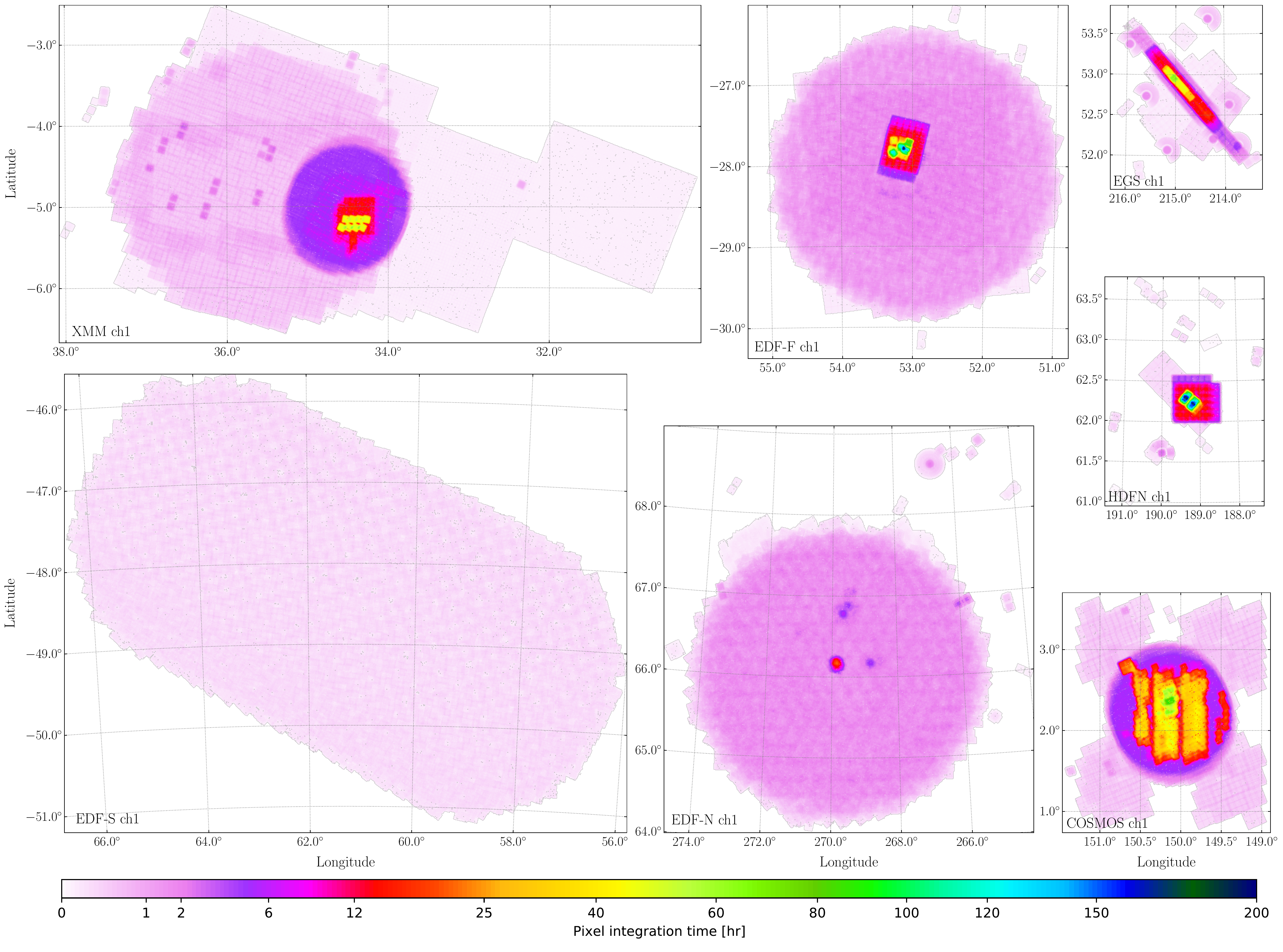}
    \caption{Integration time maps for channel 1}
    \label{fig:covs-1}
\end{figure*}

\begin{figure*}
    \centering
    \includegraphics[width=24.5cm, angle=90]{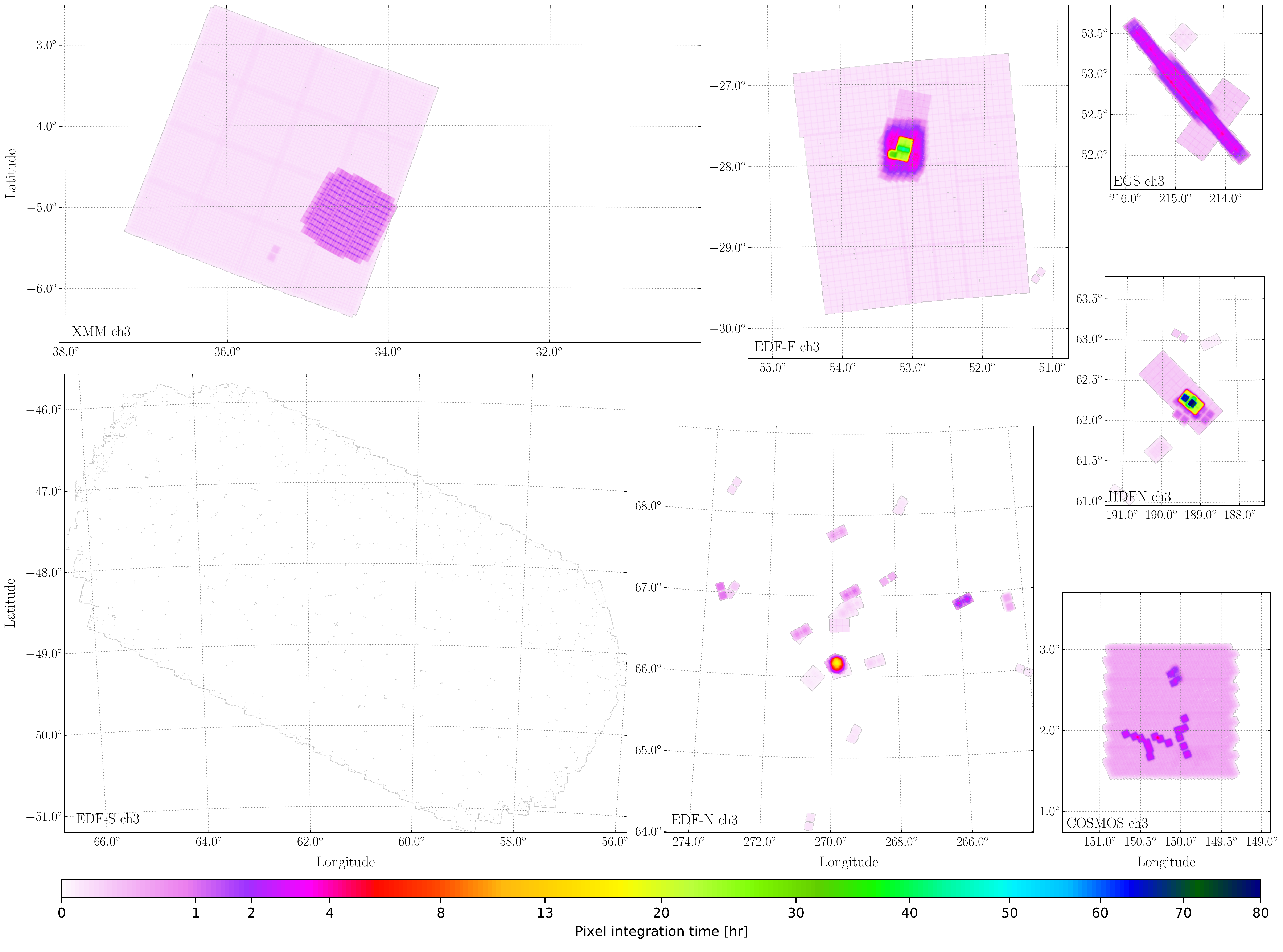}
    \caption{Integration time maps for channel 3. A blank field is included for EDF-S which was not observed in that channel and in order to have the same structure as \ref{fig:covs-1}.}
    \label{fig:covs-3}
\end{figure*}

Figures \ref{fig:covs-1} and \ref{fig:covs-3} show the full set of pixel exposure time maps for channels 1 and 3; channels, 2 and 4 are similar though slightly shifted in location. A square root scaling is applied in order to emphasise the differences at the low levels, and the same maximum is used for all fields in each channel. As EDF-S was not observed in channel 3, a blank field is placed there.  


\section{PID numbers}
\label{App:PIDs}

Table \ref{tab:pids} lists the \Spitzer Program-IDs (PIDs) of all the observations processed here. In bold the ones of the observing programmes that we planned for this work, the others are of the other archival observations that we reprocessed.
 
\begin{table*}[thb] 
\label{tab:pids}
\caption{\Spitzer Program IDs}
\begin{center}
\renewcommand{\arraystretch}{1.20} 
\setlength{\tabcolsep}{4pt}        
\begin{tabular}{l p{14cm}} \hline
Field & PIDs\\ \hline
EDF-N  & 68 609 613 618–624 1101 1125 1188 1189 1191–-1200 1317 1334 1600–-1700 1910–-1949 1951 1953--1961 1963--1983 2314 3286 3329 3672 10147 11161 \textbf{13153} 20466 30432 40385 60046 70062 70162 80109 80113 80243 80245 90209 \\ 
EDF-F & 81 82 184 194 2313 11080 \textbf{13058} 20708 30866 40058 60022 61009 61052 70039 70145 70204 80217 \\
EDF-S & \textbf{14235} \\
COSMOS & 10159 11016 12103 13094 13104 14045 14081 14203 20070 40801 50310 61043 61060 70023 80057 80062 80134 80159 90042 \\ 
EGS  & 8 10084 11065 11080 13118 20754 41023 60145 61042 80069 80156 80216 90180 \\ 
HDFN & 81 169 1304 10136 11004 11063 11080 11134 12095 13053 20218 30411 30476  40204 60122 60145 61040 61062 61063 70162 80215 \\ 
XMM &  181 3248 10042 11086 40021 60024 61041 61060 61061 70039 70062 80149 80156 80159 80218 90038 90175 90177 \\
\hline  \end{tabular} \end{center} \end{table*} 

\end{appendix}
\end{document}